\documentclass[11pt,a4paper,english]{article}
\pdfoutput=1
\usepackage{graphicx}
\usepackage{amssymb,latexsym}
\usepackage{amsmath}
\usepackage{epsf}
\usepackage{pdfsync}
\usepackage{epsfig}
\usepackage[parfill]{parskip}
\usepackage[matrix,arrow,color]{xy}
\usepackage[usenames]{color}
\usepackage{hyperref}
\usepackage{mathrsfs}
\usepackage[font={small}]{caption}
\usepackage{subcaption}
\usepackage[export]{adjustbox}
\usepackage{microtype}

\input{xy}
\xyoption{all}

\input{epsf}

\makeatletter

\usepackage{setspace}

\usepackage{lscape}

\catcode`@=11
\renewcommand{\section}
{\@startsection{section}{1}{0pt}{\medskipamount}{\medskipamount}{\large\bf}}
\makeatletter\renewcommand{\subsection}
{\@startsection{subsection}{2}{\z@}{-3.25ex plus -1ex minus -.2ex}
{1.5ex plus .2ex}{\it }}

\numberwithin{equation}{section}
\catcode`@=12
\newcommand{\ba}{\begin{eqnarray*}}
\newcommand{\ea}{\end{eqnarray*}}
\newcommand{\ban}{\begin{eqnarray}}
\newcommand{\ean}{\end{eqnarray}}

\newcommand{\tr}{{\rm tr\,}}

\newcommand{\IN}{\mathbb{N}}

\newcommand{\IK}{\mathbb{K}}

\newcommand{\cI}{{\cal I}}

\newcommand{\cN}{{\cal N}}
\newcommand{\cM}{{\cal M}}
\newcommand{\cS}{{\cal S}}
\newcommand{\cB}{{\cal B}}

\newcommand{\cH}{{\cal H}}

\newcommand{\cR}{{\cal R}}

\newcommand{\cL}{{\cal L}}
\newcommand{\cF}{{\cal F}}

\newcommand{\cK}{{\cal K}}

\newcommand{\cV}{{\cal V}}
\newcommand{\cU}{{\cal U}}
\newcommand{\cG}{{\cal G}}

\newcommand{\sfX}{{\mathsf{X}}}

\newcommand{\sfW}{{\mathsf{W}}}

\newcommand{\sfL}{{\mathsf{L}}}

\newcommand{\re}{{\rm Re \,}}
\newcommand{\im}{{\rm Im \,}}

\newcommand{\mbf}[1]{{\boldsymbol {#1} }}
\newcommand{\complex}{{\mathbb C}} %% complex numbers
\newcommand{\zed}{{\mathbb Z}} %% integers
\newcommand{\real}{{\mathbb R}} %% real numbers
\newcommand{\torus}{{\mathbb T}}

\def\ii{{\,{\rm i}\,}}
\def\dd{{\rm d}}

\newcommand{\sign}{\mathrm{sgn}}

\newcommand{\rank}{\mathrm{rank}}

\def\beq{\begin{equation}}
\def\bee{\begin{equation}}
\def\eeq{\end{equation}}
\def\bea{\begin{eqnarray}}
\def\eea{\end{eqnarray}}
\def\bd{\begin{displaymath}}
\def\ed{\end{displaymath}}

\newcommand{\Cint}{\int\kern-10.5pt-\kern7pt}

\newcommand{\PP}{{\mathbb{P}}}

\newcommand{\be}{\begin{equation}}
\newcommand{\ee}{\end{equation}}

\newcommand\fverbit{\egroup\item[\fbox{\unhbox\pippobox}]}
\newbox\pippobox

{

\def\be{\begin{equation}}
\def\ee{\end{equation}}
\def\bea{\begin{eqnarray}}
\def\eea{\end{eqnarray}}

\begin{document}

\begin{titlepage}
\setcounter{page}{1}

%\begin{flushright} IHES P/15/25 \end{flushright}

\vskip 4cm

\begin{center}

\vspace*{3cm}

{\Huge Persistent Homology and String Vacua}

\vspace{15mm}

{\large\bf Michele Cirafici}
\\[6mm]
\noindent{\em Center for Mathematical Analysis, Geometry and Dynamical Systems,\\
Instituto Superior T\'ecnico, Universidade de Lisboa, \\
Av. Rovisco Pais, 1049-001 Lisboa, Portugal}\\[4pt] 
and \\[4pt]
\noindent{\em Institut des Hautes \'Etudes Scientifiques, \\ 
Le Bois-Marie, 35 route de Chartres, F-91440 Bures-sur-Yvette, France}
\\[4pt]
Email: \ {\tt michelecirafici@gmail.com}

\vspace{15mm}

\begin{abstract}
\noindent

We use methods from topological data analysis to study the topological features of certain distributions of string vacua. Topological data analysis is a multi-scale approach used to analyze the topological features of a dataset by identifying which homological characteristics persist over a long range of scales. We apply these techniques in several contexts. We analyze $\cN=2$ vacua by focusing on certain distributions of Calabi-Yau varieties and Landau-Ginzburg models. We then turn to flux compactifications  and discuss how we can use topological data analysis to extract physical informations. Finally we apply these techniques to certain phenomenologically realistic heterotic models. We discuss the possibility of characterizing string vacua using the topological properties of their distributions.

\end{abstract}

\vspace{15mm}

\today

\end{center}
\end{titlepage} 

\newpage

\tableofcontents

% \newpage

\section{Introduction}

When studying string compactifications in many occasions one faces a set of choices among a discrete set of parameters. For instance in certain $\cN=2$ models one has to fix a Calabi-Yau variety, labelled for example by its Hodge numbers; similarly certain $\cN=1$ effective models are parametrized by the choice of a collection of integers, representing flux quanta, subject to various constraints. In all these cases we are presented with many possibilities, all of which seem equivalent before a detailed study of the low energy effective theory. There are by now many available techniques to perform such an analysis, yet one can wonder if there is a simpler setting in which one can understand qualitative features of this set of choices.

We can represent this collection of choices with a set of points. Each point could represent a particular compactification manifold, or the set of parameters determining a vacuum for a fixed background geometry. For example in a flux compactification these points could be associated with integral fluxes which stabilize the physical moduli. Or their origin could be more mysterious, for example associated with a (yet not understood) distribution of manifolds with certain properties. In the former case these points arise from studying the equations associated with the physical model, as minima of a superpotential; in the latter from some ad hoc mathematical construction, as part of the open problem of classifying higher dimensional manifolds.

In this note we pose the following question: is there any particular \textit{topological} structure in these set of points? For example one can ask if vacua in a given distribution have the tendency to cluster in distinct regions, or if the distribution of vacua presents holes or analog higher dimensional structures. Similarly one can wonder if a distribution of vacua characterized by certain physical properties is ``simple'', with almost no topological feature, or ``complex'', with many non-trivial homology generators. We will study topological features of distributions of vacua, in the appropriate sense, and consider the possibility that such topological informations could be of physical relevance. We will do so by applying techniques from topological data analysis to the problem of counting vacua in string theory. 

Topological data analysis studies how homological features of a dataset persist over a long range of scales. This is obtained by constructing a family of simplicial complexes out of a dataset and studying its homologies at various length scales. This approach to topological spaces is called \textit{persistence} \cite{ELZ02,ZC05}. The basic idea is that those homological features which are more persistent over the range of length scales can be used to give a topological characterization of the original dataset, as reviewed for example in \cite{C09,EH08,G14}. We will discuss how this characterization can be used to analyze the physical significance of a distribution of vacua; for example by measuring its topological complexity in terms of its non-trivial persistent homology classes in higher degree, or if there are certain physical requirements on the parameters which correspond to distinctive topological features.

Such techniques are becoming standard practice in data analysis, with a long range of applications, from biology and neuroscience to complex networks,  
natural images and syntax, see \cite{CIdSZ08,CCR13,jholes,NLC11,MA15} for a sample of the literature. Among the strengths of topological data analysis is its robustness respect to noisy samples, since spurious features typically show up as very short-lived persistent homology classes. For an application to the study of BPS states and enumerative problems in string and field theory see \cite{Cirafici:2015sdg}.

In plain words, the idea behind this note is simple. First of all, we take a set of string vacua, for example obtained as critical points in a flux compactification, or as a collection of compactification spaces. Out of this collection we construct a \textit{point cloud}, a set of points in $\real^N$ each representing a vacuum. Then we ``fatten" each point into a sphere of radius $\epsilon$. To this configurations we associate a continuous family of simplicial complexes by declaring that a number of points form an edge, a face or a higher dimensional simplex, if the associated spheres of radius $\epsilon$ intersect pairwise. The main idea of topological data analysis is to study such complexes \textit{as a function of $\epsilon$}. To obtain topologically invariant informations, we then pass to the homology of this continuous family of complexes, an obtain a family of homology groups parametrized by $\epsilon$. A little thought shows that the only thing that can happen is for an homology class to be born at some value $\epsilon_1$ and then to die at $\epsilon_2 > \epsilon_1$. We finally plot the lifespans (their \textit{persistence}) of all the homology classes: these are called \textit{barcodes}. At the end, we have associated a collection of barcodes to each distribution of vacua. Such a collection captures the topological features of the distribution at every length scale.

One could envision a program to study systematically string vacua from this perspective and ask which subsets of physically relevant vacua are characterized by which topological features. In this note we take a small step towards this direction by studying the persistent homology associated with a few distributions of vacua, with $\cN=2$ and $\cN=1$ supersymmetry. In this context we will learn how to interpret and extract physical information from the barcodes.

As a first step we will discuss $\cN=2$ vacua obtained from Calabi-Yau compactifications of the type II string or from Landau-Ginzburg models. Many Calabi-Yau varieties are known and have been constructed explicitly. An example is the Skarke-Kreuzer list \cite{Kreuzer:2000xy}, which parametrizes Calabi-Yau varieties obtained from reflexive polyhedra in four-dimensions. Such lists offer only samples of the full set of Calabi-Yau varieties and indeed no systematic general construction is known. From a more geometrical perspective, studying these vacua is akin to studying the topological properties of the distributions of known Calabi-Yau varieties. We will consider the question if such distributions present any particular homological feature, and if distinctive characteristics appear when restricting to geometries with certain properties, for example low Euler characteristics. We will take a similar approach to Landau-Ginzburg models.

Next we will discuss flux vacua in type IIB compactifications. As originally pointed out in \cite{Douglas:2003um,Ashok:2003gk} this is a very promising avenue to apply statistical techniques to distributions of string vacua. A great amount of work has beed dedicated to understanding the distribution of the number of flux vacua with certain properties, as reviewed for example in \cite{Douglas:2006es,Denef:2008wq,Grana:2005jc}, and part of these results have been put on firmer ground using tools from random algebraic geometry \cite{Douglas:2005df}. Here we take a topological approach, which is different from the statistical counting of vacua with certain properties and can in principle address the topological structure of the whole distribution of vacua. In this note we content ourselves with discussing the counterpart of results already known in the literature from the perspective of persistence, leaving a more detailed analysis for the future.

Finally we end with a very cursory look at a class of promising heterotic models constructed in \cite{Anderson:2011ns,Anderson:2012yf,Anderson:2013xka}. Such models are characterized by a Calabi-Yau and an holomorphic bundle, plus a series of additional choices. We will consider $\cN=1$ models which give rise to an SU(5) \textsc{gut}, which then can be higgsed by Wilson lines giving a Standard Model-like spectrum. We will study the topological features of a distribution of vacua parametrized by the Hodge numbers of the underlying Calabi-Yau and the Chern classes of the holomorphic bundle. Then we will take a somewhat different perspective, and study in a simple example how these topological features \textit{change} as we vary a certain physical parameter in the \textsc{gut} spectrum. We will use this example to make more precise the statement that a collection of vacua has a higher degree of topological complexity compared to others.

In this note we content ourselves to understand how to use techniques of computational topology to extract physical informations from distributions of string vacua and discuss which kind of facts one might hope to learn. Of course this is just a first step and one could obtain a deeper intuition by enlarging the datasets available or considering different collections of vacua. Hopefully by working with collections of larger datasets corresponding to phenomenologically viable models, one could use these methods to gain physical insights, such as if specific physical characteristics are accompanied by certain topological features. At this stage we have no evidence for this intriguing idea, and plan to investigate it more thoroughly in the future.

All the persistent homology computations in this paper have been done with \textsc{matlab} using the \textsc{javaplex} library from \cite{javaplex}. The accompanying software and dataset can be found at \cite{programs}. Manipulations of datasets have been carried out with \textsc{mathematica}.

This note is organized as follows. Section \ref{tda} gives a basic introduction to topological data analysis, presenting all the elements that we will need. In Section \ref{N2vacua} we discuss $\cN=2$ vacua obtained from Calabi-Yau compactifications and Landau-Ginzburg models. Sections \ref{flux} and \ref{heterotic} contain applications of persistent homology to the study of flux vacua in type IIB compactifications and to certain phenomenologically realistic heterotic models, respectively. Finally in Section \ref{conclusions} we summarize our findings.

%\begin{itemize}
%\item remember that you can also change the figures, for example plotting horizontally
%\item add specifics of each barcode computation: complex used, R, $\nu$, number of simplices
%\end{itemize}

\section{An introduction to topological data analysis} \label{tda}

Topological data analysis is a relatively recent approach at managing large sets of data using techniques based on computational topology \cite{ELZ02,ZC05}. It applies homological methods to collections of data arranged as point clouds to extract qualitative information. The results are only sensitive to topological information and not to geometrical quantities, such as a choice of a metric or of a system of coordinates. Furthermore it has the advantage of being functorial by construction, by studying the relations between objects as the parameters of the model are varied. The analysis of data based on topology is rapidly gaining momentum in fields such as biology, computer vision, neuroscience, languages or complex systems \cite{CIdSZ08,CCR13,jholes,NLC11,MA15}.

In this Section we will survey some basic ideas and techniques of topological data analysis. The reader interested in a more in depth discussion should consult the reviews \cite{C09,EH08,G14}, which we will follow. For a more detailed discussion aimed at physicists and various examples, see \cite{Cirafici:2015sdg}. After reviewing some basic elements of algebraic topology, we introduce the Vietoris-Rips complex and persistent homology. We also discuss some approximation schemes in the computation of barcodes and discuss how to set up the topological analysis.

\subsection{Elements of algebraic topology}

We start with a quick review of some notions of algebraic topology that we shall use in the following. In many applications it is useful to approximate a topological space by a triangulation. This can by done by means of a simplicial complex. A simplex is the convex hull of a series of points, its vertices. Its dimension is the number of vertices minus one: a 0-simplex is a single vertex, a 1-simplex is an edge between two vertices, a 2-simplex is a triangle, and so on. A face of a simplex is the convex hull of a subset of its vertices. A simplicial complex is basically a collection of simplices with the property that if a simplex is part of it, then so are all of its faces. We can state this more abstractly as follows: a \textit{simplicial complex} $S$ is a collection of non-empty sets $\Sigma$, its simplices, such that if $\sigma \in \Sigma$ and $\tau \subseteq \sigma$, then $\tau \in \Sigma$. We say that $\sigma \in \Sigma$ is a $k$-simplex if it has cardinality $k+1$. One can define maps between simplicial complexes in the natural way. A simplicial map $f$ between two simplicial complexes $S_1$ and $S_2$, is a map between the corresponding vertex sets such that a simplex $\sigma$ of $S_1$ is mapped into a simplex $f (\sigma)$ of $S_2$. A simplicial map takes a $p$-simplex into a $k$-simplex, with $k \le p$.

Let us consider a couple of examples. A simple simplicial complex which can be attached to a topological space $X$ is the nerve of a covering $\cU$, $\mathsf{Nerv} \, \cU$. Consider a covering $\cU = \{ U_i \}_{i \in I}$ labelled by a set $I$. The nerve of $\cU$ is then defined in terms of non-empty sub-collections of sets $\cS$ as
\begin{equation}
\mathsf{Nerv} \, \cU = \left\{ \cS \subseteq \cU \, \vert \, \bigcap \, \cS \, \neq 0 \right\} \, ,
\end{equation}
that is, as the simplicial complex whose vertex set is $I$ and where a set $\sigma = \{ i_0 , \cdots , i_p \} \subseteq I$ defines a simplex if and only if $U_{i_0} \cap \cdots \cap U_{i_p} \neq \emptyset$. This construction does not depend on the particular details of the covering. In particular, under general assumptions, $\mathsf{Nerv} \, \cU$ is homotopy equivalent to the space $X$. 

For a metric space $X$ consider for example the covering given by radius $\epsilon>0$ balls, $\cB_{\epsilon} (X) = \{ B_{\epsilon} (x) \}_{x \in X}$. In particular assume we can write $X = \bigcup_{i \in I} B_\epsilon (i)$ for a subset $I \subseteq X$. Then the nerve construction $\mathsf{Nerv}$ applied to the covering $\{ B_{\epsilon} (i) \}_{i \in I}$ gives the \v{C}ech complex $\mathsf{\check{C}ech}_\epsilon (I)$ associated to the set $I$ and to $\epsilon$. Note that as $\epsilon$ increases, the balls get bigger and bigger and therefore whenever $\epsilon_1 \le \epsilon_2$ we have that  $\mathsf{\check{C}ech}_{\epsilon_1} (I) \subseteq  \mathsf{\check{C}ech}_{\epsilon_2} (I)$. 

From a simplicial complex we can get interesting topological information by passing to its homology. Assume we have a simplicial complex $S$, with an ordering of the vertex set. We form the vector space of $k$-chains $C_k$ by considering linear combinations $c = \sum_i \, a_i \, \sigma_i$, where $\sigma_i$ is a $k$-simplex in $S$ and $a_i \in \zed_p$ (typically for $p$ a small prime). The boundary of a $k$-simplex $\sigma$ is the union of its $(k-1)$-subsimplices $\tau \subseteq \sigma$. One defines the boundary operator $\partial_k \, : \, C_k \longrightarrow C_{k-1}$ as
\begin{equation}
\partial_k \left( [v_0 , v_1 , \cdots , v_k] \right) = \sum_{i=0}^k (-1)^i [v_0 , \cdots , \hat{v}_i , \cdots , v_k] \, ,
\end{equation}
where the hatted variables are omitted. We can now form the chain complex
\begin{equation}
\cdots \longrightarrow C_{k+1} \longrightarrow C_{k} \longrightarrow C_{k-1} \longrightarrow \cdots
\end{equation}
and define the spaces of $k$-cycles $Z_k = \ker \, {\partial_k}$ and $k$-boundaries $B_k = \im \, {\partial_{k+1}}$. We define the homology $H_k (C_\bullet ; \zed_p)$ as the quotient $Z_k / B_k$ and its Betti numbers $b_k = \dim H_k = \dim Z_k - \dim B_k$. If the simplicial complex $S$ is derived from an underlying topological space $X$, for example via the $\mathsf{Nerv}$ or $\mathsf{\check{C}ech}$ construction, the Betti numbers give information about the topology of $X$. Heuristically $b_k$ measures the number of independent holes of dimension $k$.

Perhaps the most important feature of this construction is that it is functorial. Consider a continuous map $f : S_1 \longrightarrow S_2$ between two simplicial complexes $S_1$ and $S_2$, for example induced by a map between the two underlying topological spaces. This map induces the chain map  $C_\bullet (f) : C_\bullet (S_1) \longrightarrow C_\bullet (S_2)$ between chain complexes, such that the diagram
\begin{equation}
\xymatrix@C=8mm{  
 \cdots  \ar[r]  & C_p (S_1) \ar[r]^{\partial_p^{S_1}}  \ar[d]^{C_p (f)} & C_{p-1} (S_1) \ar[r] \ar[d]^{C_{p-1} (f)} & \cdots \\    
 \cdots  \ar[r]  & C_p (S_1) \ar[r]^{\partial_p^{S_2}} & C_{p-1} (S_1) \ar[r] & \cdots
 }
\end{equation}
commutes. The map $f$ induces the homomorphism $ f_\star : H_\bullet (S_1 ; \zed_p) \longrightarrow H_\bullet (S_2 ; \zed_p) $ at the level of the homologies, such that $f_\star [\sigma] = [f \circ \sigma]$. We will use these facts extensively in the following; in our case the map between two different simplices $S_1$ and $S_2$ will be the inclusion. Then functoriality can be used to understand the fate of the homology classes of $H_{\bullet} (S_1 ; \zed_p)$ in $H_{\bullet} (S_2 ; \zed_p)$. This idea leads to persistent homology.

\subsection{Point clouds and the Rips-Vietoris complex}

We will be interested in a version of the previous constructions. The starting point is not anymore a topological space, but a finite collection of points $\{ \mathsf{x}_i \}_{i \in I}$ in $\real^n$. We will call such a collection $\sfX$ a \textit{point cloud}. In most practical applications a point cloud is constructed out of a multidimensional data set. Topological data analysis is basically a framework which associates topological information to a point cloud, via the homology of a certain complex.

Given a point cloud $\sfX$ it is natural to define a simplicial complex whose vertices correspond to the set of points in $\sfX$. To define $k$-simplices we use a version of the $\mathsf{Nerv}$ construction. From $\sfX$ we define the space $\sfX_\epsilon = \bigcup_{\mathsf{x}_i \in \sfX} B_\epsilon (x_i)$, by fattening the points of $\sfX$. In $\sfX_\epsilon$ each point of $\sfX$ is replaced by a ball of radius $\epsilon>0$, also called the proximity parameter. For example now we can associate to $\sfX_{\epsilon}$ the \v{C}ech complex $\mathsf{\check{C}ech}_\epsilon (\sfX)$ whose vertex set is the set of points of $\sfX$ and its $k$-simplices are collections of points $\sigma = \{ \mathsf{x}_{i_0} ,  \dots , \mathsf{x}_{i_k} \}$ such that $\bigcap_{n=0}^k B_\epsilon (\mathsf{x}_{i_n}) \neq \emptyset$.

The problem with the \v{C}ech complex $\mathsf{\check{C}ech}_\epsilon (\sfX)$ is that it is computationally lengthy to check all the intersections. It is useful to approximate the \v{C}ech complex with a simpler variant, the \textit{Vietoris-Rips} complex $\mathsf{VR}_{\epsilon} (\sfX)$. To the set of points in $\sfX$ we associate the Vietoris-Rips simplicial complex as follows. Given a proximity parameter $\epsilon$, a $k$-simplex in $\mathsf{VR}_{\epsilon} (\sfX)$ is a collection of $k+1$ points $\{ \mathsf{x}_{i_0} ,  \dots , \mathsf{x}_{i_k} \}$ whose pairwise distance is less than $\epsilon$, that is
\begin{equation}
d (\mathsf{x}_i , \mathsf{x}_j) \le \epsilon \, , \qquad \text{for} \ 0 \le i,j \le \epsilon \ .
\end{equation}
Equivalently we assign balls of radius $\epsilon / 2$ to each point, and we connect two points by an edge anytime their balls intersect. The natural orientation is given by declaring that the $p$-simplex $[v_0 , \cdots , v_k]$ changes sign under an odd permutation. 

The main difference with the \v{C}ech complex $\mathsf{\check{C}ech}_\epsilon (\sfX)$, is that in defining the Vietoris-Rips complex $\mathsf{VR}_{\epsilon} (\sfX)$ we only have to compute the distance for a pair of points at a time. Furthermore the fact that in the former the parameter $\epsilon$ is the radius of the balls, while in the latter is the distance between the centers, leads to the inclusions
\begin{equation}
\mathsf{\check{C}ech}_\epsilon (\sfX) \subseteq \mathsf{VR}_{2 \epsilon} (\sfX) \subseteq \mathsf{\check{C}ech}_{2 \epsilon} (\sfX) \, .
\end{equation}
Finally now that we know how to construct the Vietoris-Rips complex $\mathsf{VR}_{\epsilon} (\sfX)$ of a point cloud $\sfX$, we can compute its homology $H_i (\mathsf{VR}_{\epsilon} (\sfX) ; \zed_p)$.

\subsection{Persistent homology and barcodes}

The idea behind persistent homology is to study the homology spaces $H_i (\mathsf{VR}_{\epsilon} (\sfX) ; \zed_p)$ as a function of $\epsilon$. Instead of a simplicial complex, now we have a collection of them, $\mathsf{VR}_{\epsilon} (\sfX)$ parametrized by $\epsilon$, which leads to a family of homology spaces $H_i (\mathsf{VR}_{\epsilon} (\sfX) ; \zed_p)$, again parametrized by $\epsilon$. While these in principle are continuous families, there will be only a finite number of inequivalent simplicial complexes which appear at finitely many $\epsilon$'s. We label these values of $\epsilon$ as $\{ \epsilon_a \}_{a \in J}$ where $J$ is a finite set.

From the point cloud $\sfX$ we construct the sequence of inclusions of spaces
\begin{equation}
\sfX_{\epsilon_0} \hookrightarrow \sfX_{\epsilon_1} \hookrightarrow \sfX_{\epsilon_2} \hookrightarrow \cdots \, ,
\end{equation}
for $0 = \epsilon_0 < \epsilon_1 < \epsilon_2 < \cdots$. For each $\sfX_{\epsilon_a}$ we can construct the associated Vietoris-Rips complex $\mathsf{VR}_{\epsilon} (\sfX)$. This leads to the filtration of simplicial complexes
\begin{equation} \label{VRfiltration}
\mathsf{VR}_{\epsilon_0} (\sfX) \hookrightarrow \mathsf{VR}_{\epsilon_1} (\sfX) \hookrightarrow \mathsf{VR}_{\epsilon_2} (\sfX) \hookrightarrow \cdots \, .
\end{equation}
Taking the $i$-th homology gives 
\begin{equation} \label{Npersistence}
H_i (\mathsf{VR}_{\epsilon_0} (\sfX) ; \zed_p)  \hookrightarrow H_i (\mathsf{VR}_{\epsilon_1} (\sfX) ; \zed_p) \hookrightarrow H_i (\mathsf{VR}_{\epsilon_2} (\sfX) ; \zed_p) \hookrightarrow \cdots \, .
\end{equation}
We will see momentarily that this is an example of an $\IN$-persistence module, and that it is completely characterized by its barcode. Let us however begin with a couple of remarks. First of all what we have defined is technically an $\real$-persistence module, since $\epsilon$ is a real variable. However only finitely many simplicial complexes will be distinct and therefore we can consider this a $\IN$-persistence module. To do this we have to chose an ordering preserving map $f \, : \, \IN \longrightarrow \real$; any choice will do, but a choice has to be made.

Secondly, it can be useful to think a bit more about \eqref{Npersistence}. The maps in \eqref{Npersistence} are the lift to homology of the inclusions between the Vietoris-Rips complexes in \eqref{VRfiltration}. Since homology is functorial,  these maps keep track of the corresponding homology classes. In other words we know if the \textit{same} homology class is present both in $H_i (\mathsf{VR}_{\epsilon_a} (\sfX) ; \zed_p)$ and $H_i (\mathsf{VR}_{\epsilon_b} (\sfX) ; \zed_p)$ for arbitrary $a$ and $b$. Therefore the only thing that can happen is for an homology class to be born at a certain ``time'' $\epsilon_a$ and then to die at a subsequent ``time'' $\epsilon_b$, with $a < b$ (we allow for the case $\epsilon_b = +\infty$). Barcodes are a simple tool to visualize homological features of a data set. They are precisely what keeps tracks of these births and deaths. %Basically one can think of barcodes as a refinement of the Betti numbers, which however extend over many homology groups. 
The idea of persistence is to look for features which persist over a large range of $\epsilon$'s. Here ``large'' can have different meaning, depending on the point cloud or on the particular questions one is asking. Persistent features are a measure of the underlying topological structure of the dataset. On the other hand, short-lived signatures are interpreted as noise, which depend on the particular approximations one is using when constructing the dataset. 

Let us make the above discussion a bit more precise. Let $\IK$ be a field. A $\IN$-\textit{persistence module} over $\IK$ is a family of vector spaces $\{ V_i \}_{i \in \IN}$ over  $\IK$, together with a collections of homomorphisms $\rho_{i,j} : V_i \longrightarrow V_j$ for every $i \le j$, such that whenever $i \le k \le j$, the homomorphisms are compatible (in the sense that $\rho_{i,k} \cdot \rho_{k,j} = \rho_{i,j}$). Persistence modules can be added, to create a new persistence module. Viceversa we can ask if a persistence module can be decomposed in simpler modules.

The usefulness of persistence modules is the existence of a classification result. This is a generalization of the similar result from elementary linear algebra, which states that finite dimensional vector spaces are classified by their dimension, up to isomorphisms. In the same fashion, certain classes of persistence modules are classified by their barcodes. We say that the persistence module $\{ V_i \}_{i \in \IN}$ is \textit{tame} if each $V_i$ is finite dimensional and $\rho_{n,n+1} : V_n \longrightarrow V_{n+1}$ is an isomorphism for large enough $n$. Given two integers $(m,n)$ so that $m \le n$, we introduce the ``interval''  $\IN$-persistence module $\IK \, (m,n)$, given by
\begin{equation}
\IK \, (m,n) = 
\begin{cases}
& 0, \qquad $if$ \ i < m \ $or$ \ i > n \\
& \IK \qquad $otherwise$
\end{cases}
\end{equation}
where $\rho_{i,j} = \mathrm{id}_\IK$ for $m \le i < j \le n$. In words $\IK (m,n)$ assigns the vector space $\IK$ to the interval $[ m , n ]$. Note that we can extend this definition for $n = + \infty$. Then the classification result states that any tame $\IN$-persistence module over $\IK$ can be decomposed as
\begin{equation}
\{ V_i \}_i \simeq \bigoplus_{j=0}^N \IK \, (m_j , n_j) \, ,
\end{equation}
for a certain $N$, and the decomposition is unique up to the ordering of the factors. In plain words a tame persistence module is completely determined by the intervals in $\IN$ where we assign non zero vector spaces. Therefore a tame $\IN$-persistence module is equivalent to the assignment of a collection of pairs of non-negative integers $(m_i , n_i)$, where $0 \le m_i \le n_i$ and we allow $n_i$ to be $+\infty$. We call such an assignment a \textit{barcode} and we represent it graphically by a collection of bars, each one associated with the aforementioned intervals.

Each collection of maps and vector spaces \eqref{Npersistence} is clearly an $\IN$-persistence module. Since the point cloud $\sfX$ is finite, it is also tame. Each $\IN$-persistence module \eqref{Npersistence}, obtained by taking homology in degree $i$, is then uniquely determined by its barcode. In the rest of this paper we will compute the barcodes associated to $\IN$-persistence modules which arise from certain point clouds and discuss the physical interpretation of their persistent features.

% We have been very brief in our discussion of persistent homology. A slightly more detailed review aimed at physicists is in \cite{Cirafici:2015sdg}

\subsection{Approximation schemes and witness complexes}

When a point cloud $\sfX$ consists of a large number of points, the computation of the Vietoris-Rips simplicial complexes and of the associated $\IN$-persistence modules can become quite intractable. We will now discuss certain approximation schemes, based on the idea that one could select only a limited subset $\sfL$ of $\sfX$ as vertices of the simplicial complexes.

A landmark selector is an operator which chooses the subset $\sfL$ from $\sfX$. Each landmark selector has its own advantages and disadvantages and it is important to be aware of them. An obvious landmark selector picks the elements of $\sfL$ at random. This is quite useful, although depending on how scattered is the dataset, it may miss important features. We will mostly use a so-called \textit{maxmin} selector. The idea is to select points by induction,  in order to maximize the distance from the already chosen set. More precisely we start from a randomly chosen point. The remaining ones are chosen by induction: if $\sfL_i$ consists of $i$ chosen landmark points, then the next $i+1$-th landmark point is chosen in order to maximize the function $\mathsf{z} \longrightarrow d (\sfL_i , \mathsf{z})$, where $d(\sfL_i , \mathsf{z})$ is the distance between the landmark set $\sfL_i$ and the point $\mathsf{z} \in \sfX$. Note that with a maxmin landmark selection, the landmark set will consist of points spreading apart from each other as much as possible. As a consequence the set will cover the dataset, in principle better than a random selection. The drawback is that the maxmin algorithm will generically choose outlier points. 

A landmark selector can greatly simplify the computation of the homology. We will use landmark selection to define approximations to the Vietoris-Rips complex, the witness complexes. Assume we have chosen a landmark set $\sfL$ from a point cloud $\sfX$. We define the witness complex $\mathsf{W} (\sfX,\sfL,\epsilon)$ as follows. The vertex set of $\sfW (\sfX,\sfL,\epsilon)$ is given by $\sfL$. To define simplexes, we pick a point $\mathsf{x} \in \sfX$ and we denote by $m_k (\mathsf{x})$ the distance between $\mathsf{x}$ and its $k+1$-th closest landmark point. Then a collection of $k > 0$ vertices $\mathsf{l}_i$ form a $k$-simplex $[ \mathsf{l}_0  \dots  \mathsf{l}_k ]$ if all its faces are in $\sfW (\sfX,\sfL,\epsilon)$ and there exists a witness point $\mathsf{x} \in \sfX$ so that
\begin{equation}
\max \{ d (\mathsf{l}_0 , \mathsf{x}) , d (\mathsf{l}_1 , \mathsf{x}) , \dots , d (\mathsf{l}_k , \mathsf{x}) \} \le m_k (\mathsf{x}) + \epsilon \, .
\end{equation}
In particular we have the inclusion $\sfW (\sfX , \sfL , \epsilon_1) \subseteq \sfW (\sfX , \sfL , \epsilon_2)$ when $\epsilon_1 < \epsilon_2$. Therefore we can construct a filtration of witness complexes as function of the proximity parameter $\epsilon.$ Passing to the homology defines $\IN$-persistence modules and we can study their persistent features by looking at the barcodes.

Another approximation scheme is the lazy witness complex $\mathsf{LW} (\sfX , \sfL , \epsilon)$. Again this complex depends on a landmark selection, but this time an extra parameter $\nu \in \mathbb{N}$ is involved. The vertex set of $\mathsf{LW} (\sfX,\sfL,\epsilon)$ is again given by $\sfL$. To define simplexes we need to introduce a notion of distance. For $\mathsf{x} \in \sfX$ we define $m(\mathsf{x})$ as the distance between $\mathsf{x}$ and the $\nu$-th closest landmark point if $\nu$ is non zero, and set $m(\mathsf{x})=0$ otherwise. Then, given two vertices $\mathsf{l}_1$ and $\mathsf{l}_2$ in $\sfL$, the edge $[\mathsf{l}_1 \mathsf{l}_2]$ is in $\mathsf{LW} (\sfX , \sfL , \epsilon)$ if we can find a witness point $\mathsf{x} \in \sfX$ so that
\begin{equation}
\max \{ d (\mathsf{l}_1 , \mathsf{x}) , d (\mathsf{l}_2 , \mathsf{x})  \} \le m (\mathsf{x}) + \epsilon \ .
\end{equation}
Finally a higher dimensional simplex is an element of the lazy witness complex $\mathsf{LW} (\sfX , \sfL , \epsilon)$ if all of its edges are.  Note that again the inclusion $\mathsf{LW} (\sfX , \sfL , \epsilon_1) \subset \mathsf{LW} (\sfX , \sfL , \epsilon_2)$ holds whenever $\epsilon_1 < \epsilon_2$ and again we can construct filtrations by inclusion. The usefulness of the lazy witness complex is that it is less computationally involved since it is determined from its 1-skeleton. The lazy witness complex is the simplest way to study the persistent homology of a dataset. In this paper we will always set $\nu=1$.

In this paper we will use \textsc{matlab} to perform all the homology computations and our programs will employ the library \textsc{javaplex}, available from \cite{javaplex}. We will use \textsc{mathematica} for the manipulations of the datasets. Our \textsc{matlab} programs and datasets are available at \cite{programs}.

\subsection{Topological analysis}

Finally we collect some qualitative ideas on how the topological analysis based on persistent homology will be used in our context. Again we refer the reader to \cite{C09,EH08,G14} for more examples of the practical uses of these techniques.
\begin{enumerate}
\item Topological data analysis provides qualitative information about a dataset. Heuristically it determines the topological properties of a dataset, such as its clustering in connected components, or the presence of loops or in general higher dimensional surfaces. Being topological, this information is independent on the set of coordinates or any metric used for the analysis. For example by regarding a dataset as a statistical approximation to an underlying topological manifold at a certain length scale, one computes the Betti numbers of such a manifold. From the barcodes one obtains information about the homologically non-trivial $n$-cycles as well as their characteristic length scale as measured by the proximity parameter $\epsilon$.
\item The presence of barcodes in higher degree indicates that the data are organized forming higher dimensional homologically non-trivial cycles, at least at a certain length scale. This can be seen as evidence of existing correlations between the data. For example, if in a certain region of the point cloud the points are disposed along an $n$-cycle, it is possible that there exists relations between themselves, in the form of a series of algebraic equations which describe the $n$-cycle. 
\item Short-lived persistent homology classes are generically regarded as noise, while long-lived classes point out towards homologically robust features. This is intuitively clear and follows from the definition of the Vietoris-Rips complex and its variations. Therefore one is lead to look for long-lived bars, encoding the persistent homology classes. Of course, what does it mean to be short- or long-lived is somewhat subjective and depends sensitively on the physical problem. We will see example of physically interesting but short-lived persistent homology classes. In particular this can happen when the existence of a symmetry or modular property forces the same behavior at various length scales.
\item The topological analysis can be most effective when comparing different datasets. This perspective is commonly used in fields such as biology or neuroscience, where the barcodes of different datasets can reveal if a certain drug was effective or not. In our case one can for example select string vacua with or without a certain feature, say a certain particle present in the low energy spectrum, and ask if this comports qualitative differences, and of which type. 
\end{enumerate}
In the following we will apply these ideas on certain distributions of string vacua.

\section{$\cN=2$ vacua} \label{N2vacua}

%In the previous Section we have assumed a fixed compactification manifold and worked out the topological properties of the distribution of flux vacua. In principle this information is already contained in the superpotential for each choice of fluxes and one could imagine that it could be derived explicitly given stronger analytical tools. One the other hand there is not such a function which gives Calabi-Yau manifolds. 

Having set up our main computational tools, we proceed to use them in a few specific examples of $\cN=2$ vacua of the type II string. Examples of such vacua include Calabi-Yau manifolds and Landau-Ginzburg models. The construction of Calabi-Yaus is an art on its own and while hundreds of thousands of examples are known explicitly, the list is far from exhaustive. The classification problem is still wide open, and the origin of Calabi-Yau varieties still rather mysterious. It is natural to wonder if the techniques we have exposed so far can be applied to the known distributions of Calabi-Yau varieties, and what kind of information can we hope to gain. Similar arguments hold in the case of Landau-Ginzburg models, which play an important part in the classification of $\cN=2$ superconformal field theories.

In particular we will study the following (incomplete) set of string vacua:
\begin{itemize}
\item The Skarke-Keuzer list from \cite{KSlist}, containing Calabi-Yaus which can be realized as a hypersurface in a toric variety, and correspond to four dimensional reflexive polyhedra. Of this list, 30108 varieties have distinct Hodge numbers.
\item Complete Intersection Calabi-Yaus (CICYs), which are constructed via a complete intersection of polynomials within a product of projective spaces. We take the list from \cite{CICYlist}, which contains the 7890 CICYs constructed in \cite{Candelas:1987kf}. From this list we take the 266 pairs of distinct Hodge numbers and add their mirrors.
\item A list of Landau-Ginzburg models, their abelian orbifolds and certain models with discrete torsion, taken from \cite{LGlist}. We parametrize these models by $(\chi, n+\bar{n})$ (for simplicity we remove by hand all those models which need an extra label, which are characterized by $\chi=0$). 
\end{itemize}

\subsection{Calabi-Yau compactifications}

We begin by considering vacua of the type II string which have the form $\real^{3,1} \times X$ where $X$ is a Calabi-Yau threefold, a complex three dimensional manifold with trivial canonical bundle. These compactifications preserve $\cN=2$ supersymmetry in $\real^{3,1}$. The Calabi-Yau theorem states that for each K\"ahler class $\omega \in H^{1,1} (X ; \complex)$ there exists a unique Ricci flat K\"ahler metric. The moduli space of Calabi-Yau metrics is parametrized by $h^{2,1} (X) = \dim  H^{2,1} (X ; \complex)$ complex structure deformations and $h^{1,1} (X) = \dim  H^{1,1} (X ; \complex)$ K\"ahler moduli, which correspond to scalar fields in the effective four dimensional theory. These Hodge numbers characterize the low energy effective action by determining geometrically the number of vector multiplets and hypermultiplets. The Euler characteristic of a Calabi-Yau is given by $\chi (X) = 2 (h^{1,1} (X) - h^{2,1} (X))$.

The superconformal theories which describe the propagation of strings on Calabi-Yaus come in pairs, a phenomenon known as mirror symmetry. Mirror symmetry has been established for many pairs of Calabi-Yaus. In this case we say that two Calabi-Yaus form a mirror pair $(X,Y)$ and they represent the same physical vacuum. The Hodge numbers for mirror pairs are related as $h^{1,1} (X) = h^{2,1} (Y)$ and $h^{2,1} (X) = h^{1,1} (Y)$. More deeply complexified K\"ahler moduli and complex structure moduli are exchanged. Often certain quantities associated with a Calabi-Yau can be computed exactly and this leads to interesting mathematical predictions for the mirror manifold. Such an example are quantum correction due to worldsheet instantons which modify the low energy effective action and are associated with an extremely interesting enumerative problem, Gromov-Witten theory, counting holomorphic curves on $X$. Gromov-Witten theory produces symplectic invariants of $X$. In this note we will only consider the cruder topological information contained in the Hodge numbers and Euler characteristic; applications of persistent homology to enumerative problems of Calabi-Yau appear in \cite{Cirafici:2015sdg}.

Therefore as a first approximation we label a Calabi-Yau compactification by its Hodge numbers and Euler characteristic as $(h^{1,1} (X)+h^{1,2} (X), \chi (X))$. Of course this is a rather crude approximation, since different Calabi-Yaus can have the same Hodge numbers. We would like to consider a distribution of these values as a point cloud $\sfX$, where a Calabi-Yau $X$ corresponds to the point $ \mathsf{x}= (h^{1,1} (X)+h^{1,2} (X), \chi (X))$, and study its persistent homology. A related question is if point clouds $\sfX$ obtained in this way from collections of varieties arising from different constructions have different topological features, as seen from their barcodes. We will address these questions with certain known lists of Hodge numbers of Calabi-Yaus.

The simplest construction of a Calabi-Yau manifold is as a hypersurface in a complex projective space. For example the quintic threefold can be seen as the vanishing locus of a homogenous polynomial of degree five $f_5 (z_1 , \dots , z_5) = 0$, where $[ z_1 , \dots , z_5 ]$ denote the homogeneous coordinates of $\PP^4$. The K\"ahler form $\omega$ of the quintic descends from the K\"ahler form of $\PP^4$, while the complex structure moduli correspond to the independent deformations of the defining equation $f_5 (z_1 , \dots , z_5) = 0$. More general constructions can be thought of as more elaborate versions of this simple example and many lists of Calabi-Yau families are available in the literature. 

An example of such a list is the class of complete intersection Calabi-Yaus (CICYs), that is those which can be constructed via the complete intersection of polynomials in a product of projective spaces. Their classification was completed in \cite{Candelas:1987kf,Candelas:1987du} and their Hodge numbers computed in \cite{Green:1987cr}. Such a list is available at \cite{CICYlist} (to which we add the mirror Hodge numbers). Another list was obtained by Kreuzer and Skarke in \cite{Kreuzer:2000xy}, by classifying reflexive polyhedra in four-dimensions. Reflexive polyhedra in four-dimensions describe Calabi-Yau threefolds by realizing them as hypersurfaces in toric varieties \cite{Batyrev:1994hm}, and the classification proceeds using the powerful combinatorial techniques of toric geometry.

%\begin{figure}[htbp]
%\centering
%\def\svgwidth{0.4cm}
%\includegraphics[width=0.5\textwidth]{KSList}
%\caption{KSList}
%\label{KSList}
%\end{figure} 
%
%\begin{figure}[htbp]
%\centering
%\def\svgwidth{0.4cm}
%\includegraphics[width=0.5\textwidth]{CICYList}
%\caption{CICYList}
%\label{CICYList}
%\end{figure} 

\begin{figure}[htbp]
\centering
\begin{minipage}[b]{0.47\textwidth}
\centering
\def\svgwidth{0.47cm}
\includegraphics[width=1.1\textwidth]{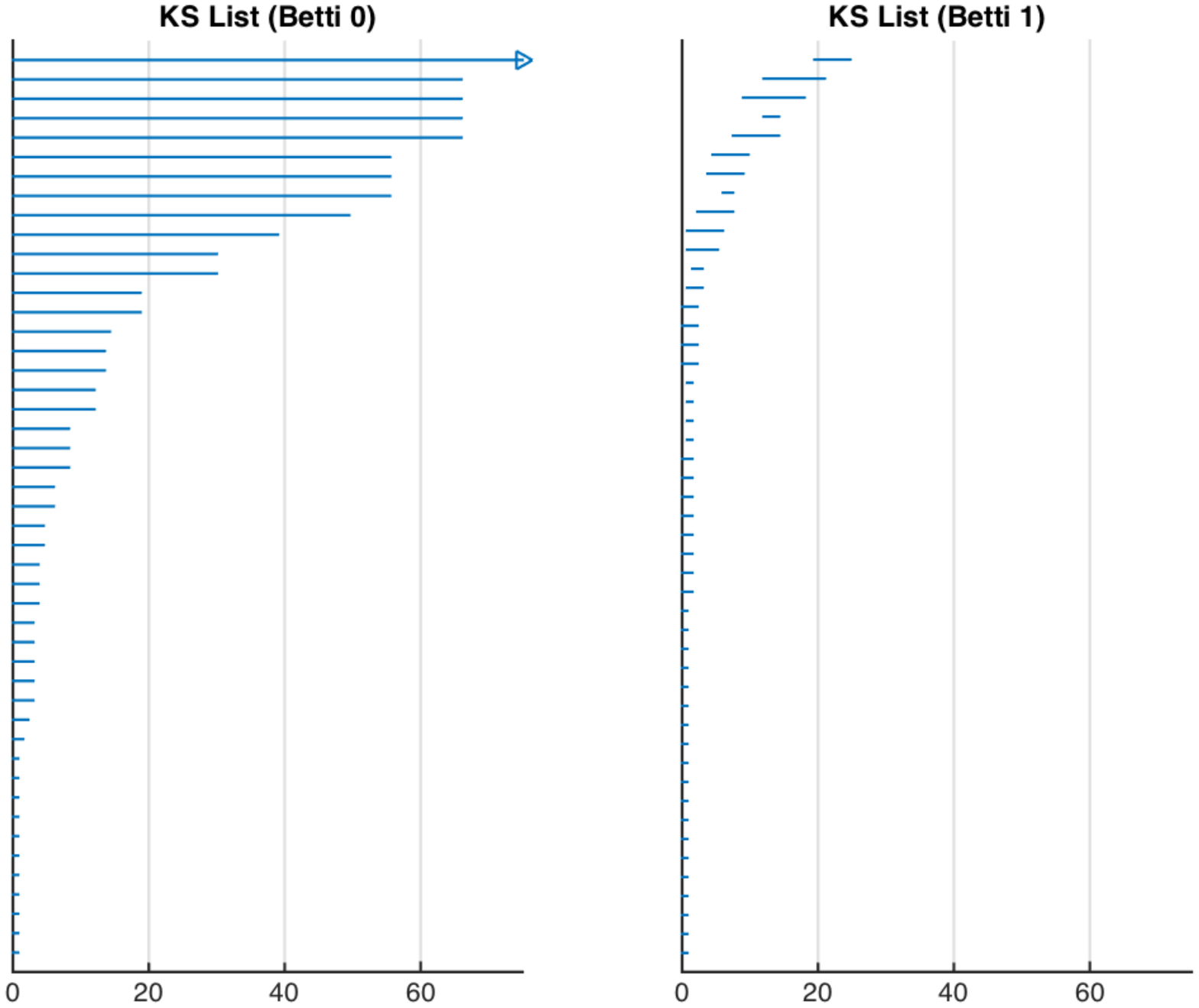}
%\caption{KSList}
%\label{KSList}
\end{minipage} 
\qquad
%\hspace{0.5cm}
\begin{minipage}[b]{0.47\textwidth}
\centering
\def\svgwidth{0.47cm}
\includegraphics[width=1.1\textwidth]{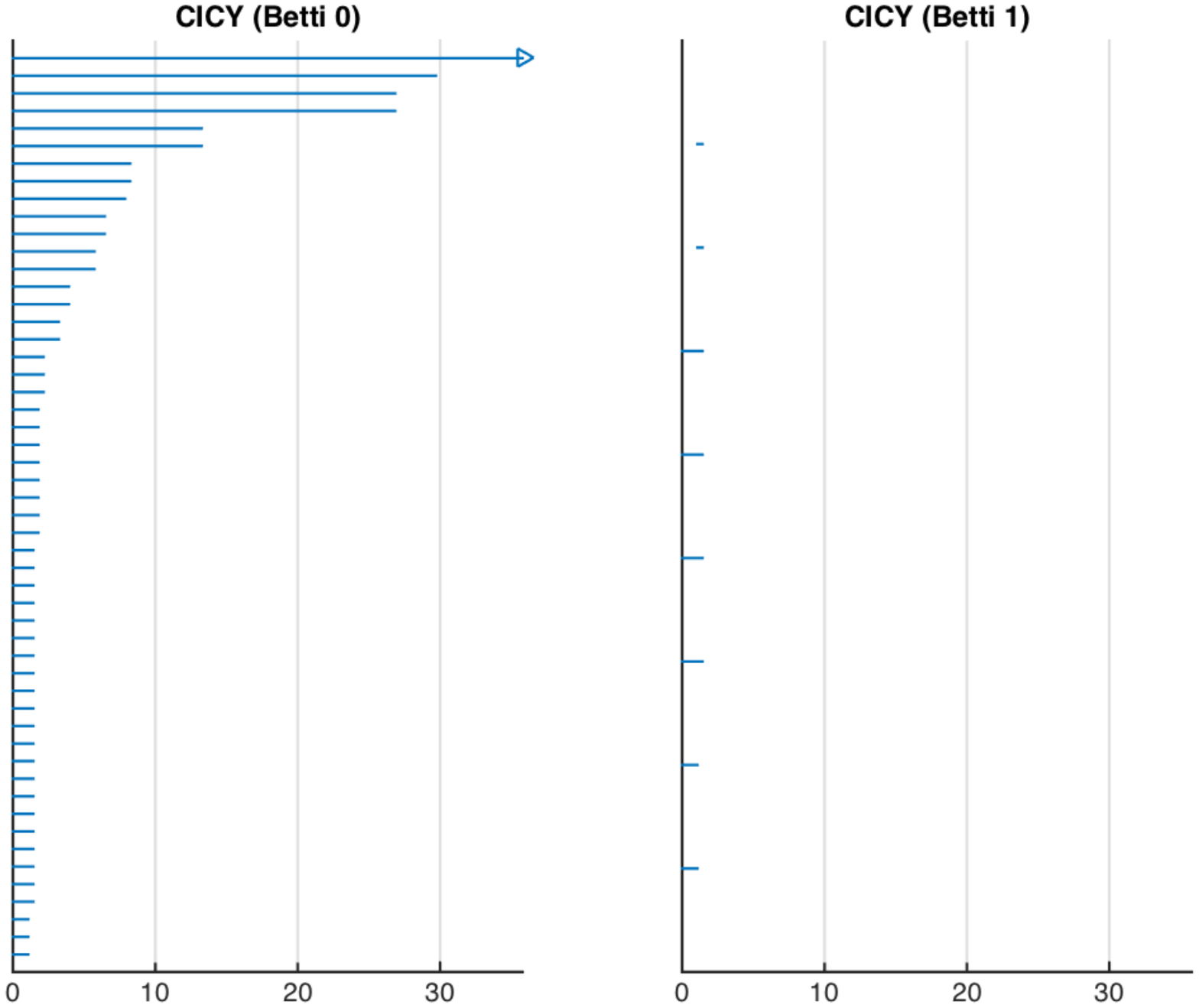}
%\caption{CICYList}
%\label{CICYList}
\end{minipage}
\caption{Barcode computation for the distribution of Calabi-Yau manifolds. The point cloud $\sfX$ is constructed assigning to a Calabi-Yau $X$ the vector $ \mathsf{x}= (h^{1,1} (X)+h^{1,2} (X), \chi (X))$. The computation of the persistent homology is done using the lazy witness complex $\mathsf{LW (X,L,\epsilon)}$. The landmark set consists of 200 points in both cases. \textit{Left.} The Kreuzer-Skarke list of Calabi-Yaus. The point cloud consists of 30108 points; the homology computations runs over 49744 simplices. \textit{Right.} Complete Intersection Calabi-Yaus, with their mirrors. The point cloud has 532 elements, and the homology computation involves  251926  simplices.}
\label{KSList}
\label{CICYList}
\end{figure} 

Out of each one of these two lists we construct a point cloud and study its persistent homology using the lazy witness complex $\mathsf{LW (X,L,\epsilon)}$. To do so we wrote a program in \textsc{matlab}, available at \cite{programs}, using the library \textsc{javaplex}. The results are shown in Figure \ref{KSList}. On the left we have the barcodes corresponding to the modules $H_{\bullet} (\mathsf{LW (X , L , \epsilon)} ; \zed_2)$ for the Kreuzer-Skarke list of Calabi-Yaus corresponding to reflexive polyhedra, on the right the list of CICYs to which we have added the mirror Hodge numbers. Recall that bars corresponding to $\IN$-persistence modules in degree zero (we will informally call these ``Betti number 0'') are a measure of the number of connected components, at every length scale. Similarly a barcode in degree one is a signature of non-trivial 1-cycles. The collection of varieties under consideration does not have any barcode in higher degree. The appearance of 1-cycles in Figure \ref{KSList} on the left, measured by the barcode in degree one, means that there are zones  where no Calabi-Yau is present\footnote{Of course one should also take into account that the Hodge numbers are integers while the proximity parameter is a real variable. This will induce a few spurious 1-cycles, which are however too small to be noticed in Figure \ref{KSList}.}, in the list we are considering. The empty regions in the distribution of Calabi-Yaus are detected at values of $\epsilon$ for which the points surrounding these areas are closer than $\epsilon$ (at this value we see the birth of an homology class) and disappear at values of $\epsilon$ greater than the characteristic length of the empty region, which is now filled up (and the homology class dies because it becomes a boundary).

Most bars in the barcode in degree zero are relatively short-lived. This is a sign that the distribution of manifolds is generically rather dense, since at small values of $\epsilon$ nearby points become part of the same simplex, ceasing to be isolated connected components. For visible values of $\epsilon$ in Figure \ref{KSList} (\textit{left}) the vast majority of connected components associated with individual points has already disappeared. This is not true for every bar, pointing out to the existence of isolated points or clusters in the distribution. An interesting feature of Figure \ref{KSList} is that certain bars come in pairs: this is a consequence of mirror symmetry, or more down to earth the symmetry of the Hodge numbers. The mirror bars correspond to two areas of the distribution with exactly the same behavior. Of course this is not true for any bars: areas which are close to the symmetry axis in the Hodge numbers distribution will start to ``interact'' with each other as $\epsilon$ increases before showing any mirror structure, becoming a single connected component.

Of course all these features could equivalently be ``seen'' by plotting the Hodge number distribution. The purpose of this discussion is to turn them into a  mathematically precise statement concerning its topological features. The formalism of $\IN$-persistence modules provides the necessary tools.

A similar discussion holds for the distribution of CICYs, in Figure \ref{CICYList} (\textit{right}). Note that all sign of non-trivial homological structures disappear at values $\epsilon \sim 30$ of the proximity parameter, much smaller than in the case of the KS list. Now the barcode in degree one shows even less structure, a sign that the CICYs are more evenly distributed and the formation of 1-cycles is accidental. This is an example of short-lived homology classes which can be regarded as ``noise'' and excluded from the persistence analysis\footnote{We have confirmed this conclusion by repeating the computation for different landmark selections.}. From this perspective the set of CICYs is topologically simpler than the set of varieties in the KS list.

Now we focus on a particular zone, namely the ``tip'' of the distribution, that is the region with small $h^{1,1} (X)+h^{2,1} (X)$. This region was identified as  special in \cite{Candelas:2007ac}, on the ground that heterotic models with small Hodge numbers engineer supersymmetric extensions of the Standard Model with few generations of fermions. Such Calabi-Yaus appear to be rather rare and the corresponding area quite unpopulated. We wish to examine this area from the point of view of persistent homology. To begin with we collect all the manifolds in the KS and CICY lists, and add the other small Hodge number Calabi-Yaus pointed out in \cite[Table 1]{Candelas:2007ac}, and their mirrors. We stress again that we are parametrizing Calabi-Yaus by their Hodge numbers, which is a very crude approximation since distinct varieties have the same Hodge numbers.
\begin{figure}[htbp]
\centering
\begin{minipage}[b]{0.47\textwidth}
\centering
\def\svgwidth{0.47cm}
\includegraphics[width=1.1\textwidth]{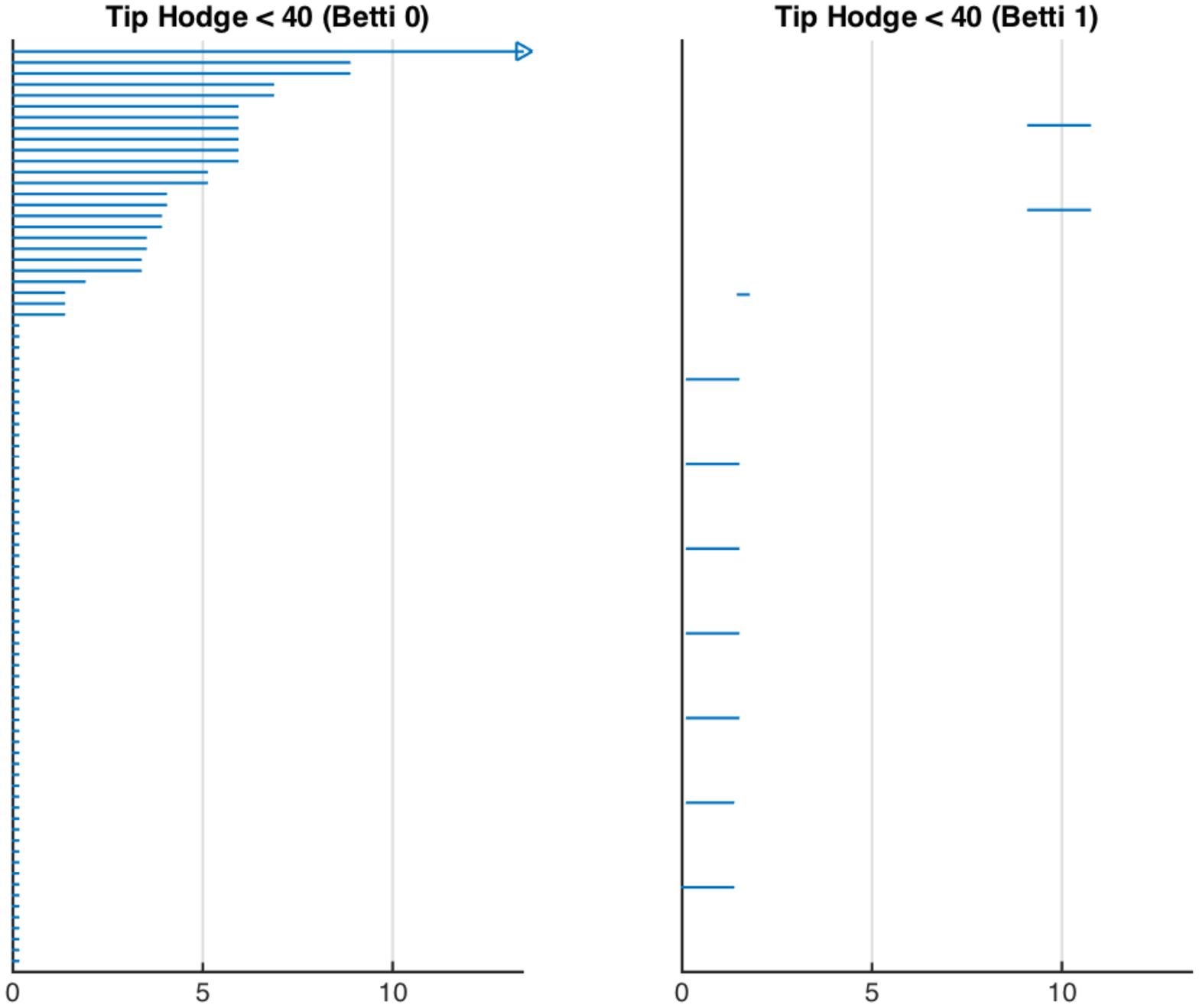}
%\caption{CYTipListLazy}
%\label{CYTipListLazy}
\end{minipage} 
\qquad
%\hspace{0.5cm}
\begin{minipage}[b]{0.47\textwidth}
\centering
\def\svgwidth{0.47cm}
\includegraphics[height=0.8\textwidth,left]{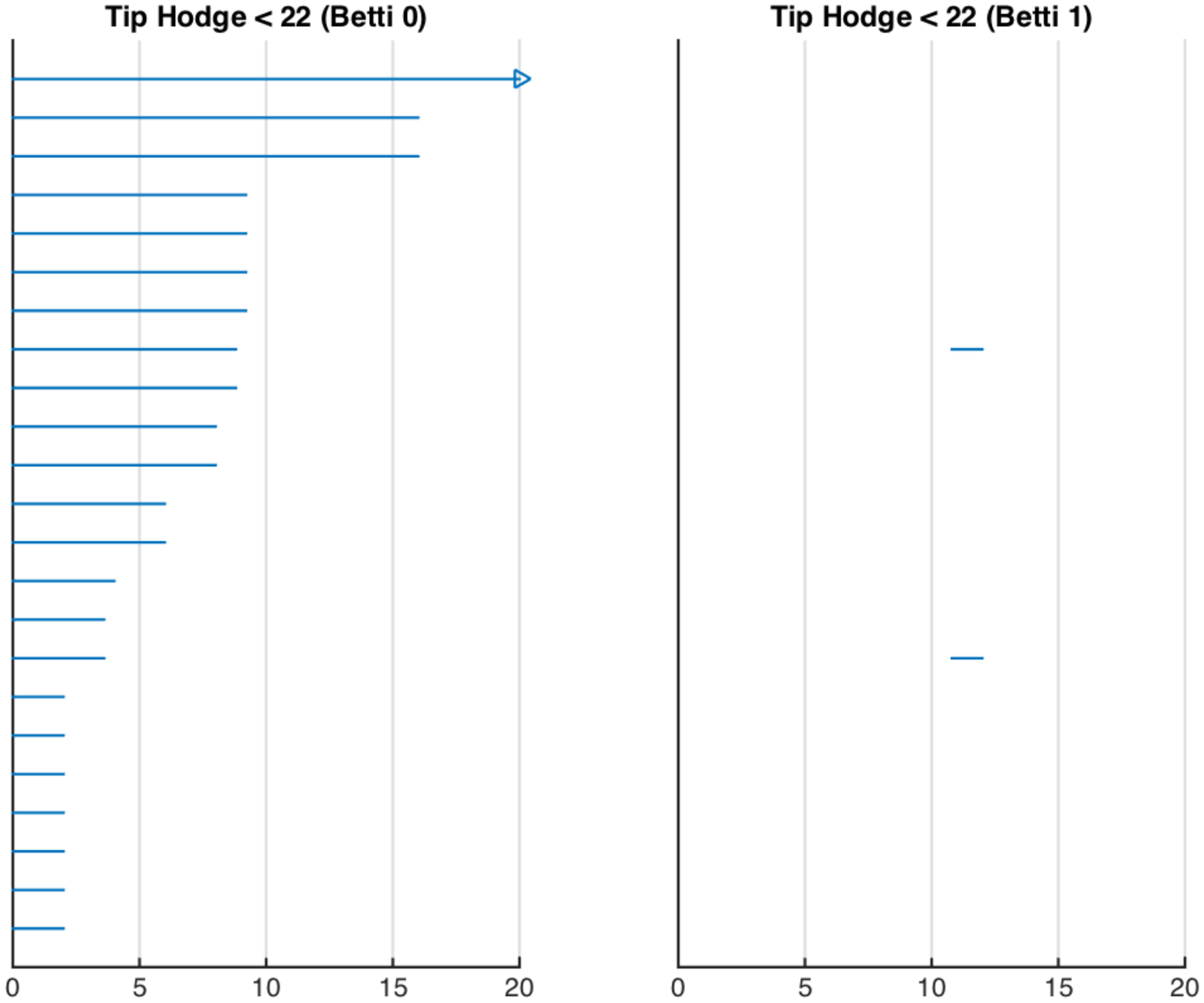}
%\caption{CYTipOnly}
%\label{CYTipOnly}
\end{minipage}
\caption{Barcode computation for Calabi-Yau varieties with small Hodge numbers. The point cloud is constructed as in Figure \ref{KSList}. \textit{Left.} The point cloud parametrizes Calabi-Yaus with $h^{1,1} (X)+h^{2,1} (X) < 40$ and consists 479 points. The persistent homology is computed via the lazy witness complex $\mathsf{LW (X, L ,\epsilon)}$ over $\zed_2$ with a landmark selection $\mathsf{L}$ of 170 points, and run over 118362 simplices. \textit{Right.} The point cloud parametrizes 23 Calabi-Yaus with $h^{1,1} (X)+h^{2,1} (X) < 22$. The filtered homology computation employs the Vietoris-Rips complex $\mathsf{VR}_\epsilon (X)$ over $\zed_2$ and run overs 827 simplices.}
\label{CYTipListLazy}
\label{CYTipOnly}\end{figure} 
In Figure \ref{CYTipListLazy} we show the barcodes for said distributions, with Hodge numbers $h^{1,1} (X)+h^{2,1} (X) < 40$ (\textit{left}) and $h^{1,1} (X)+h^{2,1} (X) < 22$ (\textit{right}). Let us consider the Figure on the left. We see that in degree one two bars stand out with respect to the others, signaling the formation of two 1-cycles at a larger length scale. Such cycles are larger than the rest, as they appear at larger values of $\epsilon$. 

We can try to give a topological rephrasing of the philosophy of \cite{Candelas:2007ac}, according to which Calabi-Yau varieties with small Hodge numbers are ``special''. In our language this would translate into the statement that the distribution of Calabi-Yaus with small values of the Hodge numbers has certain distinctive topological features. Indeed these specific 1-cycles continue to exist if we restrict our point cloud to Calabi-Yaus with $h^{1,1} (X)+h^{2,1} (X) < 22$, as shown in Figure \ref{CYTipOnly} on the right. The two odd features at Betti number one appear clearly, even if rather short-lived. The fact that they come in a pair is a consequence of mirror symmetry.

We have seen an instance of one of the main themes of this paper, that special vacua are associated with characteristic topological features. Clearly the analysis so far has been rather limited; however we think it elucidates the principle that one might expect that topological interesting structures at the level of persistent homology correspond to physical interesting settings.

\subsection{Landau-Ginzburg vacua}

Now we turn to Landau-Ginzuburg vacua. Such models describe two dimensional $\cN = 2$ superconformal field theories. They are governed by a number of chiral superfields $\Phi_i$, $i=1,\dots,N$ interacting via a quasi-homogeneous superpotential $W (\Phi_i)$. We will assume that $W (\Phi_i)$ has isolated and non-degenerate critical points only. Ground states are related to elements of the chiral ring $\cR$, obtained by taking the quotient of the space spanned by the chiral superfields by the Jacobian ideal of the superpotential $W (\Phi_i)$
\begin{equation}
\cR = \frac{\complex[\Phi_1 , \dots , \Phi_N]}{\langle \partial W (\Phi_i) \rangle} \, .
\end{equation}
The degeneracies of chiral primaries are encoded into the Poincar\'e polynomial
\begin{equation} \label{Poinpol}
P(t,\overline{t}) = \tr_{\cR} t^{J_0} \overline{t}^{\overline{J}_0} = \sum_{ij} \, p_{ij} t^i \bar{t}^j \, ,
\end{equation}
expressed in terms of generators of the superconformal algebra. We will also consider only Landau-Ginzburg models with central charge $c=9$. Certain orbifolds of $\cN=2$ Landau-Ginzburg models describe conformal field theories at certain points of the Calabi-Yau moduli spaces. Such models can also be used in heterotic string compactifications, where they engineer an effective four dimensional theory with $E_6$ gauge symmetry and matter in the $\mathbf{27}$ representation. The fermionic spectrum is determined by the chiral ring $\cR$ from \eqref{Poinpol}, namely $p_{11} = n_{\mathbf{27}}$ is the number of fermion in the $\mathbf{27}$ representation and $p_{12} = n_{\overline{\mathbf{27}}}$. If the Landau-Ginzburg model has a geometrical interpretation as strings propagating in a Calabi-Yau $X$, then we can identify the Hodge numbers $h^{1,1}(X) = p_{12}$ and $h^{2,1}(X) = p_{11}$. We will assume that the other non-trivial coefficients of \eqref{Poinpol} vanish, if necessary discarding the respective models from the lists of consistent vacua.

We consider the collection of 10839 models constructed in \cite{Kreuzer:1992np} by the classification of non-degenerate quasi-homogeneous polynomials which can play the role of a superpotential of a $c=9$ model. The massless spectra can be computed from \eqref{Poinpol} and result in 2997 different spectra (or pairs of Hodge numbers). This list was extended in \cite{Kreuzer:1992iq} by classifying all the abelian symmetries of the above potentials which can be used to construct abelian orbifolds. This results in 3798 inequivalent spectra, which we will use for our analysis. An additional list of models can be obtained by considering discrete torsion as in \cite{Kreuzer:1994qp}, which results in 138 extra spectra.
\begin{figure}[htbp]
\centering
\begin{minipage}[b]{0.47\textwidth}
\centering
\def\svgwidth{0.47cm}
\includegraphics[width=1.1\textwidth]{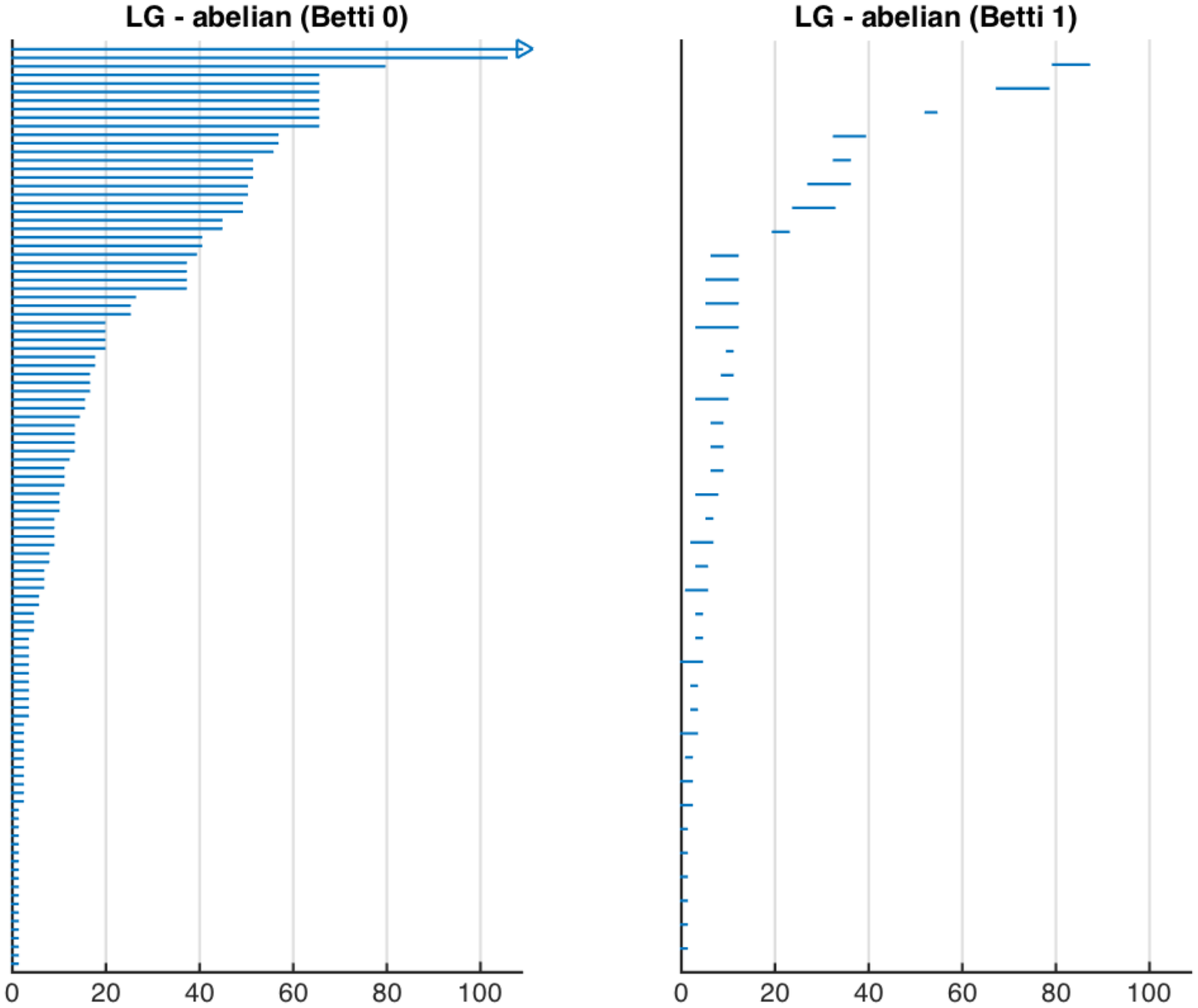}
%\caption{LGab}
%\label{LGab}
\end{minipage} 
\qquad
%\hspace{0.5cm}
\begin{minipage}[b]{0.47\textwidth}
\centering
\def\svgwidth{0.47cm}
\includegraphics[width=1.1\textwidth]{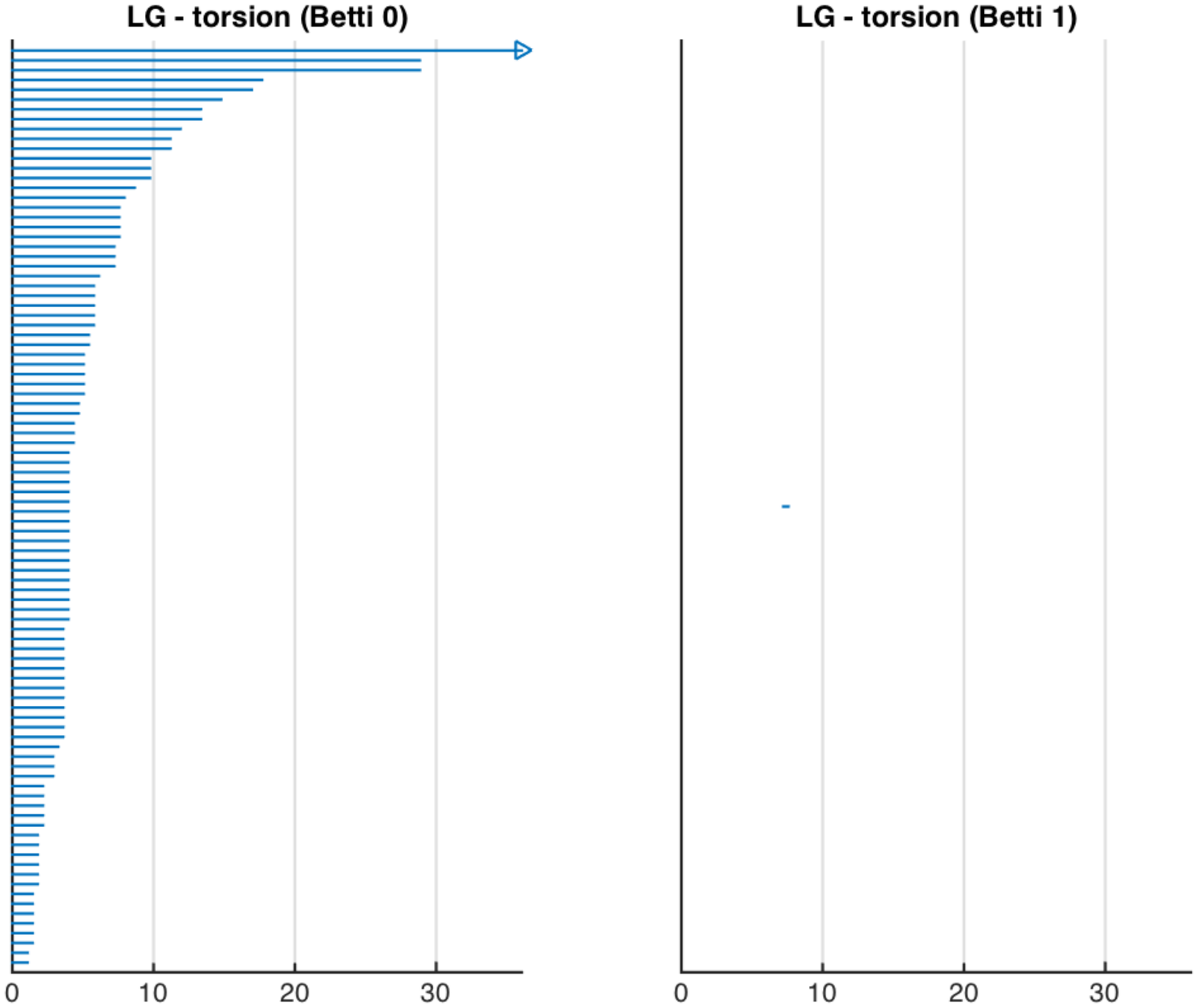}
%\caption{LGtorsLazy}
%\label{LGtorsLazy}
\end{minipage}
\caption{Barcodes for Landau-Ginzburg models. The point clouds $\sfX$ are constructed out of vectors of the form $\mathsf{x} = (\chi = 2 (p_{12}-p_{11}) , p_{11}+p_{12})$. To pass to the homology we use the lazy witness complex $\mathsf{LW} (\sfX , \mathsf{L} , \epsilon)$ over $\zed_2$. \textit{Left.} Abelian orbifolds. The point cloud $\sfX$ consists of 3792 points and the landmark set $\mathsf{L}$ of 200. The computation of filtered homology requires 123057 simplices. \textit{Right}. Discrete torsion models. The point cloud $\sfX$ parametrizes 128 models, 100 of which constitute the landmark set $\mathsf{L}$. The homology computation runs over 57090 simplices.}
\label{LGtorsLazy}
\label{LGab}
\end{figure}
We have studied the persistent homology for the distributions of abelian orbifolds and discrete torsion models, taking as input the set of inequivalent spectra. We have labeled a Landau-Ginzburg spectrum by the vectors $\mathsf{x} = (\chi = 2 (p_{12}-p_{11}) , p_{11}+p_{12})$ as explained above, and used these to define our point cloud $\sfX$. We have then computed the associated homology groups using the lazy witness complex $\mathsf{LW} (\sfX , \mathsf{L} , \epsilon)$ as a function of $\epsilon$.

The barcodes resulting from the homology computation for $H_i \left( \mathsf{LW} (\sfX , \mathsf{L} , \epsilon) ; \zed_2 \right)$ are shown in Figure \ref{LGab}. The distribution of abelian orbifold vacua, on the left, appears to be the most complex from a topological viewpoint, also with respect to Figures \ref{KSList} and \ref{CYTipOnly}. This is most apparent in degree one, where several 1-cycles appear with no obvious regularity. In degree zero the Figure shows the existence of many long-lived connected components, a signal that the vacua tend to cluster in certain regions. On the other hand the distribution of discrete torsion models is extremely simple, with no features in degree one.

From this perspective the two type of vacua appear topologically very different, despite arising from similar constructions. In this sense the topological analysis can discriminate between two physically similar situations. 

\section{Persistence in flux vacua} \label{flux}

Now we proceed to apply our techniques to flux vacua of the type II\,B string. These are realized by an orientifold of a Calabi-Yau compactification, with fluxes turned on along compact cycles and a collection of D-branes. We will only consider complex structure and axion-dilaton moduli, stabilized by the flux induced superpotential. We are not really interested (yet) in fully stabilized and phenomenologically viable models, but we wish to analyze simple examples to show how to use techniques of topological data analysis and which kind of results one can expect.

%Are certain properties (ie chiral matter, number of generations) associated with any particular topological feature.

\subsection{Flux compactifications}

Before discussing the uses of persistent homology, we briefly review some basic elements of flux compactifications. We will consider II\,B/F-theory vacua obtained by an orientifold of a Calabi-Yau compactification, as reviewed in \cite{Douglas:2006es,Denef:2008wq,Grana:2005jc}. In general one can use D3 branes, extended in the directions transverse to $X$, and $D7$ branes, wrapping holomorphic cycles to obtain a quasi-realistic $\cN=1$ model where moduli are stabilized by fluxes and quantum corrections. We will be mostly interested in the distributions of flux vacua, and ignore issues of phenomenological viability of the model or of the simplifying assumptions. 

We take a Calabi-Yau $X$, with $h^{2,1} (X)$ complex structure deformations and $h^{1,1} (X)$ K\"ahler moduli. For simplicity we will fix the K\"ahler class, so that Calabi-Yau metrics are parametrized by complex structure deformations. The relevant moduli space parametrizing physical configurations is non-compact and has the form $\cM = \cM_c \times \mathcal{H} / SL(2 ; \zed)$, where $\cM_c$ is the complex structure moduli space and $\cH$ is the upper half plane where the axion-dilaton $\tau$ takes values. The axion-dilaton $\tau = C_0 + \ii / g_s$ is a function of the Ramond-Ramond scalar $C_0$ and the string coupling constant $g_s$. Equivalently such a space can be interpreted in F-theory as the moduli space of Calabi-Yau metrics on $X \times \torus^2$ where the $\torus^2$ is elliptically fibered over $X$.

Let $\{ A_a , B^a \}$ with $a,b=1 , \dots , h^{2,1}+1$ be a symplectic basis of $H_3 (X , \real)$ and $\{ \alpha_a , \beta^a\}$ its Poincar\'e dual integral cohomology basis, so that
\begin{equation} \label{symplecticbasis}
\int_{A_a} \, \alpha_b = \delta_b^a \, , \qquad \int_{B^b} \, \beta^a = - \delta^a_b \, , \qquad \int_X \, \alpha_a \wedge \beta^b = \delta_a^b \ .
\end{equation}
A choice of the complex structure $z \in \cM_c$ determines the Hodge decomposition
\begin{equation}
H^3 (X , \complex) = H^{3,0}_z (X) \oplus  H^{2,1}_z (X) \oplus  H^{1,2}_z (X) \oplus  H^{0,3}_z (X) \, .
\end{equation}
On a Calabi-Yau $H^{3,0}_z (X)$ is one dimensional and the fibration $H^{3,0}_z (X) \longrightarrow \cM_c$ defines the Hodge bundle. Take $\Omega_z \in H^{3,0}_z (X)$. Similar arguments hold for the axion-dilaton modulus, so the actual Hodge bundle of physical interest here is $H^{3,0}_z (X) \otimes H_\tau^{1,0} (\torus^2) \longrightarrow \cM$, where we write $\omega_\tau \in H^{1,0}_\tau (\torus^2)$ as $\omega_\tau = \dd x + \tau \, \dd y$.

In our basis the $(3,0)$ form $\Omega_z$ for $z \in \cM_c$ can be represented as $\Omega_z = z^a \, \alpha_a - \cG_a \, \beta^a$ in terms of its periods
\begin{equation}
\int_{A_a} \, \Omega_z = z^a \, , \qquad \int_{B^a} \, \Omega_z = \cG_a (z) \, ,
\end{equation}
where $z^a$ are local projective coordinates on the complex structure moduli space, and the functions $\cG_a (z)$ can be expressed as the derivatives of a single function, the prepotential $\cG(z)$, as $\cG_a (z) = \partial_a \, \cG(z)$. The period vector $\Pi (z) = (\cG_b (z) , z^a)$ plays an important role in the determination of the K\"ahler potential and flux superpotential. The Hodge bundle has a natural hermitian metric, the Weil-Petersson metric and the associated K\"ahler potential on $\cM$ is
\begin{align}
\cK (z , \tau) &= - \log \left( \ii \int_X \Omega \wedge \bar{\Omega} \right) - \log \left( - \ii (\tau - \bar \tau) \right) \cr
& =  - \log \left( - \ii \Pi^\dagger \cdot \Sigma \cdot \Pi \right) - \log \left( - \ii (\tau - \bar \tau) \right) \, ,
\end{align}
where and $\Sigma$ is the symplectic matrix of rank $2 h^{2,1}+2$.

Now we turn on fluxes in the NS and RR field strengths $F_3 , H_3 \in H^3 (X , \zed)$. Since fluxes are quantized, they  can be expressed in terms of the integer valued vectors $\mbf f$ and $\mbf h$ as
\begin{align}
F_3 &= - (2 \pi)^2 \alpha' \left( f_a \, \alpha^a + f_{a + h^{2,1}+1} \, \beta_a \right) \, , \\
H_3 &= - (2 \pi)^2 \alpha' \left( h_a \, \alpha^a + h_{a + h^{2,1}+1} \, \beta_a \right) \, .
\end{align}
with $a=1 , \dots , h^{2,1}(X)+1$. We assemble these fluxes in the 4-form on $X \times \torus^2$ given by $G_3 = F_3 \wedge \dd y + H_3 \wedge \dd x$.

The presence of non-trivial fluxes generates a non-trivial superpotential \cite{Gukov:1999ya} which stabilizes complex structure moduli. Such a superpotential is a section of the line bundle $\cL$ dual to the Hodge bundle, given by the functional
\begin{align}
W_G (z , \tau) &= \int_{X \times \torus^2} G_3 \wedge \Omega_z \wedge \omega_\tau \\
&= \int_X  \left( F_3 - \tau \, H_3 \right) \wedge \Omega_z = (2 \pi)^2 \alpha' \left( {\mbf f} - \tau \, {\mbf h} \right) \cdot \Pi (z)
\end{align}
acting on sections $\Omega_z \wedge \omega_\tau$ of the Hodge bundle.

In local coordinates the F-term equations which follow from the superpotential $W_G (z , \tau)$ are
\begin{align} \label{EOMtau}
D_\tau W_G (z, \tau) &=  \partial_\tau W_G (z,\tau)  + W_G (z,\tau) \, \partial_\tau \, \cK (z , \tau)  \cr &=  (2 \pi)^2 \alpha' \, ({\mbf f} - \bar \tau \, {\mbf h}) \cdot \Pi (z) = 0 \, , \\
D_{z^a} W_G (z , \tau) & = \partial_{z^a} W_G (z,\tau) + W_G (z,\tau) \, \partial_{z^a} \, \cK (z , \tau)  \cr &=  (2 \pi)^2 \alpha' \, ({\mbf f} - \tau \, {\mbf h}) \cdot (\partial_{z^a} \, \Pi (z) + \Pi (z)  \, \partial_{z^a} \cK (z , \tau) ) = 0 \, , \label{EOMcompl}
\end{align}
where $D$ is the connection associated with the Weil-Petersson metric on $\cM$. The critical point equations \eqref{EOMtau} and \eqref{EOMcompl} define supersymmetric vacua.

Turning on fluxes gives a contribution to the total $D3$ brane charge
\begin{equation}
N_{\rm flux} = \frac{1}{(2 \pi)^4 (\alpha')^2} \int_X F_3 \wedge H_3 = {\mbf f} \cdot \Sigma \cdot {\mbf h} \, . 
\end{equation}
If we denote by $N_{D3}$ the number of $D3$ branes transverse to $X$, consistency of the compactification requires the tadpole cancellation condition
\begin{equation} \label{tadpole}
N_{\rm flux} + N_{D3} = L \, .
\end{equation}
In our cases, the orientifold can be seen as arising from an F-theory compactification. This involves a four-fold $Z$ which is an elliptic fibration over $X / g$, with $g$ the orientifold involution. In this case $L = \chi(Z) / 24$. In particular this sets a bound $N_{\rm flux} \le L$ on the total amount of flux available. Therefore to find supersymmetric critical points, one has to solve the equations \eqref{EOMcompl} in a certain region of the moduli space, as the fluxes take values satisfying the tadpole constraint.

Compactifications of this sort have been widely studied since they provide workable models where ideas about the statistical distributions of string vacua can be tested \cite{Douglas:2003um,Ashok:2003gk,Denef:2004ze}. We will have nothing to say about this approach, but we mention a few points for completeness. A simple distribution of string vacua is the density
\begin{equation} \label{dens}
\dd \mu (z) = \sum_i \, \delta_z \left(  D_i W_G (z,\tau) \right) \, ,
\end{equation}
which is just a sum of delta function contributions, each one centered at a supersymmetric vacuum (here $i$ runs over axion-dilaton and complex moduli and we are again neglecting K\"ahler moduli). This distribution is in general pretty intractable and often it is simpler to deal with an approximate continuum distribution $\rho (z)$. The integral of such a distribution over a certain region $\mathcal{U}$ of the moduli space corresponds to the number of supersymmetric vacua satisfying certain properties, which enter in the definition of $\rho (z)$. A slightly different approach is to define an index density $\dd \, \cI$ which differs from \eqref{dens} only in that each delta function is weighted by the sign of the Jacobian $(-1)^F = \sign \det_{i,j} D^2 W_G (z,\tau)$. One motivation for doing so is that such an index appears to be the proper generalization of the Morse index for dealing with vacua in supergravity. Indeed while in rigid supersymmetry the total number of vacua is often given by a topological formula, in supergravity vacua can be created or destroyed in pairs, precisely the situation of Morse theory. This analogy, discussed more in depth in \cite{Denef:2004ze},  suggests that techniques based on persistent homology, which can be thought of as a version of Morse theory, could be usefully applied to the statistical distributions of vacua. In this paper we will focus on actual solution of the F-term equations, and leave the analysis of statistical distributions to the future. An example of such a index density is \cite{Ashok:2003gk}
\begin{equation} \label{indexvacua}
\cI \sim \int_{\cM} \, c_{\rm top} \left( T^* \cM \otimes \cL \right) \, ,
\end{equation}
the top Chern class of the bundle where $D \, W_G (z,\tau)$ takes values. Physically such an integral gives an estimate of the number of supersymmetric vacua. One can refine such a formula by counting vacua with specific properties, say a certain value of the cosmological constant. We will not consider such counting problems in this paper. However it is interesting to note that the number of vacua from \eqref{indexvacua} is almost a topological quantity (it would be for compact $\cM$).

The point which is important for us is that a high degree of topological complexity of the moduli space of physical configurations $\cM$ or of the dual Hodge bundle $\cL$, corresponds to a large number of critical points for the superpotential $W_G (z , \tau)$ and therefore to a large number of supersymmetric vacua. In other words we expect a distribution of vacua to have a high degree of topological complexity. Persistent homology is precisely a tool which can measure the topological complexity of a distribution of points. We therefore expect to obtain interesting informations by applying techniques of statistical topology to ensembles of string vacua. In the remaining of this Section we will see examples of such applications in simple models.

\subsection{Rigid Calabi-Yau}

We begin with the simplest case, a rigid Calabi-Yau with no complex structure moduli, as studied in \cite{Ashok:2003gk}. Having no complex structure moduli, $h^{2,1}$ vanishes and therefore $b_3=2 + 2 h^{2,1} = 2$. This implies that the periods of the holomorphic three form $\Omega$ over the symplectic basis $\{ A , B \}$ of \eqref{symplecticbasis} can be taken to be $\Pi = (1 , \ii)$. Therefore the flux superpotential is
\begin{equation}
W_G (\tau) = \left( -h_1 - \ii h_2 \right) \, \tau + (f_1 + \ii f_2) \, ,
\end{equation}
where the fluxes vectors  ${\mbf f} = (f_1,f_2)$ and ${\mbf h = (h_1 , h_2)}$ take values in  $H^3 (X , \zed)$ and are constrained by the tadpole cancellation condition $f_1 \, h_2 - h_1 \, f_2 \le L$. The F-term equation $D_\tau \, W_G (\tau) = 0$ has the simple solution
\begin{equation}
\tau = \frac{f_2 + \ii f_1}{h_2 + \ii h_2} \ .
\end{equation}
The moduli space is the upper half plane $\cH$ modulo the action of $\rm{SL} (2 , \zed)$. To solve this equation we use \textsc{mathematica} to generate random vectors of integral fluxes obeying the tadpole constraint and only retain those values of $\tau$ which lie in a fundamental domain $\cF$ for $\rm{SL} (2 , \zed)$ in $\cH$. This restriction is to avoid overcounting of physically equivalent vacua which lie in the same $\rm{SL} (2 , \zed)$ orbit.

\begin{figure}[htbp]
\centering
\def\svgwidth{0.4cm}
\includegraphics[width=0.7\textwidth]{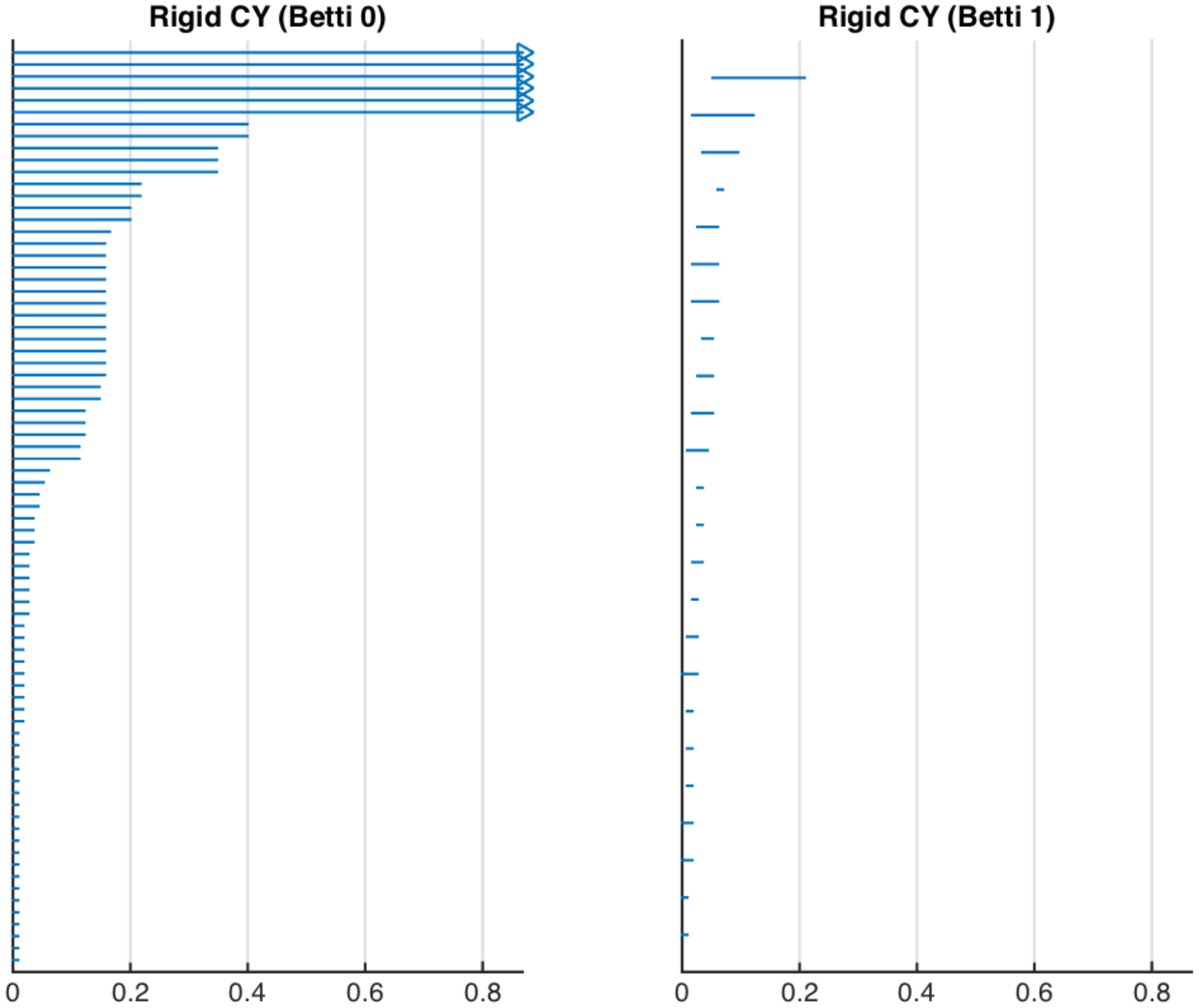}
\caption{Barcodes for the distribution of flux vacua in a rigid Calabi-Yau. The point cloud $\mathsf{X}$ consists of 1064958 inequivalent vacua. We compute the persistent homology using the lazy witness complex $\mathsf{LW} (\mathsf{X} , \mathsf{L} , \epsilon)$ with a landmark set $\mathsf{L}$ consisting of 140 points. The computation runs over 203362 simplices.}
\label{TauRigidCY}
\end{figure}

Then we use the values $\mathsf{x} = (\re \tau , \im \tau)$ to construct a point cloud $\mathsf{X}$ and study its persistent homology via the lazy witness  complex $\mathsf{LW} (\mathsf{X} , \mathsf{L} , \epsilon)$. The corresponding barcodes are shown in Figure \ref{TauRigidCY}.
From the distribution of the barcodes we learn the following facts. The structure of the barcode with Betti number one imply the presence of relatively long-lived and short-lived regularly distributed 1-cycles in the point cloud. These corresponds to the ``voids'' in the distribution of vacua, already noted in \cite{Ashok:2003gk}: the distribution contains holes, at the center of which there is a big degeneracy of vacua. The presence of these holes implies that at certain values of the filtration parameter $\epsilon$, non-trivial 1-cycles will form in the filtered homology. These cycles have different sizes, some are smaller (short-lived bars) and other are larger (long-lived bars). The presence of several degree one bars which are born and die at the same value of $\epsilon$ implies that the corresponding cycles have roughly the same size. The long-lived bars correspond to bigger cycles, since it takes more $\epsilon$ time to cover them up.

The presence of a big degeneracy of vacua at the center of these holes shows up as a connected component in the degree zero barcode. Such bars are the very short-lived classes in Figure \ref{TauRigidCY}. This identification follows from the fact that a number of these connected components disappear from the barcode's plot roughly at the same $\epsilon$ time as the 1-cycles. This is an example of what we can learn not just by looking at the barcodes by themselves but also at the correlations between the barcodes in different degrees. Note that such a perspective is typically not common in topological data analysis, where the very short-lived bars in the barcode in degree zero would have been interpreted as noise. We must be very careful in applying such techniques to string theory vacua since due to the action of the mapping class group on the moduli space $\cM$ it is possible that interesting physical features are mapped to short-lived homology classes in the fundamental domain. This mechanism is important and worth stressing. The $\rm SL(2 , \complex)$ action is a symmetry which relates physically equivalent configurations. To avoid overcounting one has to restrict the attention to a fundamental domain $\mathcal{F}$, which contains precisely a single representative for each $\rm SL(2, \complex)$ orbit. However nothing guarantees that topological interesting features which are prominent in a fundamental domain, will remain so in another. One way in which we can avoid misidentifying these features is to  cross correlate the barcodes' plots at different degrees, at least in our example. Most importantly we have now learned how to recognize such features.

On the other hand the formation of long-lived bars in degree zero, which implies the existence of more connected components at larger values of $\epsilon$ is a consequence of the data set being limited. This is a sign that the flux vacua are not uniformly distributed but tend to accumulate at small values of $\im \tau$. Since the data set is limited, points at higher values of $\im \tau$ will be more rare when the sample of vacua is generated at random, and therefore will appear as long lived connected components (uncorrelated with any structure at Betti number one). 

All of these conclusions could easily be drawn just by looking directly at the plot of flux vacua on the upper half plane, see for example \cite[Figure 6]{Denef:2004ze}. We have discussed this dataset in detail as an example of how these features would show up from the perspective of the barcodes. Now we turn to higher dimensional point clouds for which a simple plot is not available.

\subsection{A Calabi-Yau hypersurface example}

Now we turn to a more sophisticated example, studied in \cite{Giryavets:2003vd,Giryavets:2004zr,DeWolfe:2004ns}, where the moduli space $\cM_c$ parametrizing complex structure deformations is one dimensional. We will consider the Calabi-Yau $X_8 (1^4,4)$, defined as the hypersurface
\begin{equation}\label{hypersurM8}
\sum_{i=1}^4 x_i^8+4 \, x_0^2-8 \, \psi \, x_0 \, x_1 \, x_2 \, x_3 \, x_4 = 0
\end{equation}
in the weighted projective space $\PP^4 (1^4,4)$ (where the notation $w^m$ denotes the $m$ times repetition of the weight $w$). The relevant Hodge numbers of $X_8 (1^4,4)$ are $h^{1,1} (X_8 (1^4,4)) = 1$ and $h^{2,1} (X_8 (1^4,4)) = 149$. The defining equation \eqref{hypersurM8} is invariant under the acton of the discrete group $\Gamma= \zed_8^2 \times \zed$. The variable $\psi$ parametrizes complex structure deformations of the mirror $\widetilde{X}$, which according to the Greene-Plesser orbifold construction \cite{Greene:1990ud} is a crepant resolution of $X / \Gamma$. The complex structure moduli space of $\widetilde{X}$ can be identified with $\PP^1 \backslash \{0,1,\infty \}$, where the three special points correspond to a Landau-Ginzburg point, a conifold point and the large complex structure point respectively. 

In general complex structure moduli of $X_8 (1^4,4)$ are not invariant under $\Gamma$. If we restrict our attention to the periods which are invariant under the $\Gamma$ action, the corresponding Picard-Fuchs equations simplify greatly. One is left with a reduced period vector $\tilde{\Pi} = (\cG_1 , \cG_2 , z_1 , z_2)$ which as a first approximation is not a function of the remaining moduli: those can only appear as higher order monomials which are invariant under the $\Gamma$ action. This period vector coincides with the period vector of $\tilde{X}$  \cite{Giryavets:2003vd}.

We are interested in an orientifold of this model which breaks supersymmetry down to $\cN=1$. The orientifold in question acts as $x_0 \longrightarrow - x_0$ and $\psi \longrightarrow - \psi$, as well as worldsheet parity reversal. We turn on flux vectors $\mbf h$ and $\mbf f$ taking values in $H_3 (X_8 (1^4,4) , \zed)$ and compatible with the $\Gamma$ action, along the 3-cycles associated with the period vector $\tilde{\Pi}$. The tadpole cancellation condition \eqref{tadpole} is now
\begin{equation}
N_{D3} + N_{\rm flux}  = 972 \, .
\end{equation}
Complex structure moduli which transform non-trivially under $\Gamma$ can appear in the superpotential only at higher order as $\Gamma$-invariant monomials and will be stabilized at zero, the fixed point of $\Gamma$ \cite{Giryavets:2003vd}. They will be neglected in the following. To summarize the only relevant moduli for the model at hand are the complex structure modulus $\psi$ and the axion-dilaton $\tau$.

To be more concrete, we will also consider a particular region of the moduli space, nearby the conifold point $\psi_{\rm con} = 1$. It will be useful to introduce an auxiliary variable $x = 1-\psi$ which measures the distance from the conifold locus. In this region the period vector can be explicitly computed as a function of $x$ and we shall use the result of \cite{Giryavets:2004zr,DeWolfe:2004ns}. Explicitly the solutions of equations \eqref{EOMtau} and \eqref{EOMcompl} have the form
\begin{align} \label{EOMtauM8}
\tau &= \frac{f_1 \overline{a}_0 +  f_2 \overline{b}_0 +f_3 \overline{c}_0 }{h_1 \overline{a}_0 +  h_2 \overline{b}_0 +h_3 \overline{c}_0 } + \cdots \, , \\
\log x & = -\frac{2 \pi \ii}{d_1 (f_2 - \tau \, h_2)} \, \sum_{i=1}^4 \,  \left( f_i - \tau \, h_i \right) \, A_i \, ,  \label{EOMcomplM8}
\end{align}
in terms of a set of known constants, which the reader can find in \cite[Section 5]{Giryavets:2004zr}.
\begin{figure}[htbp]
\centering
\def\svgwidth{0.4cm}
\includegraphics[width=0.7\textwidth]{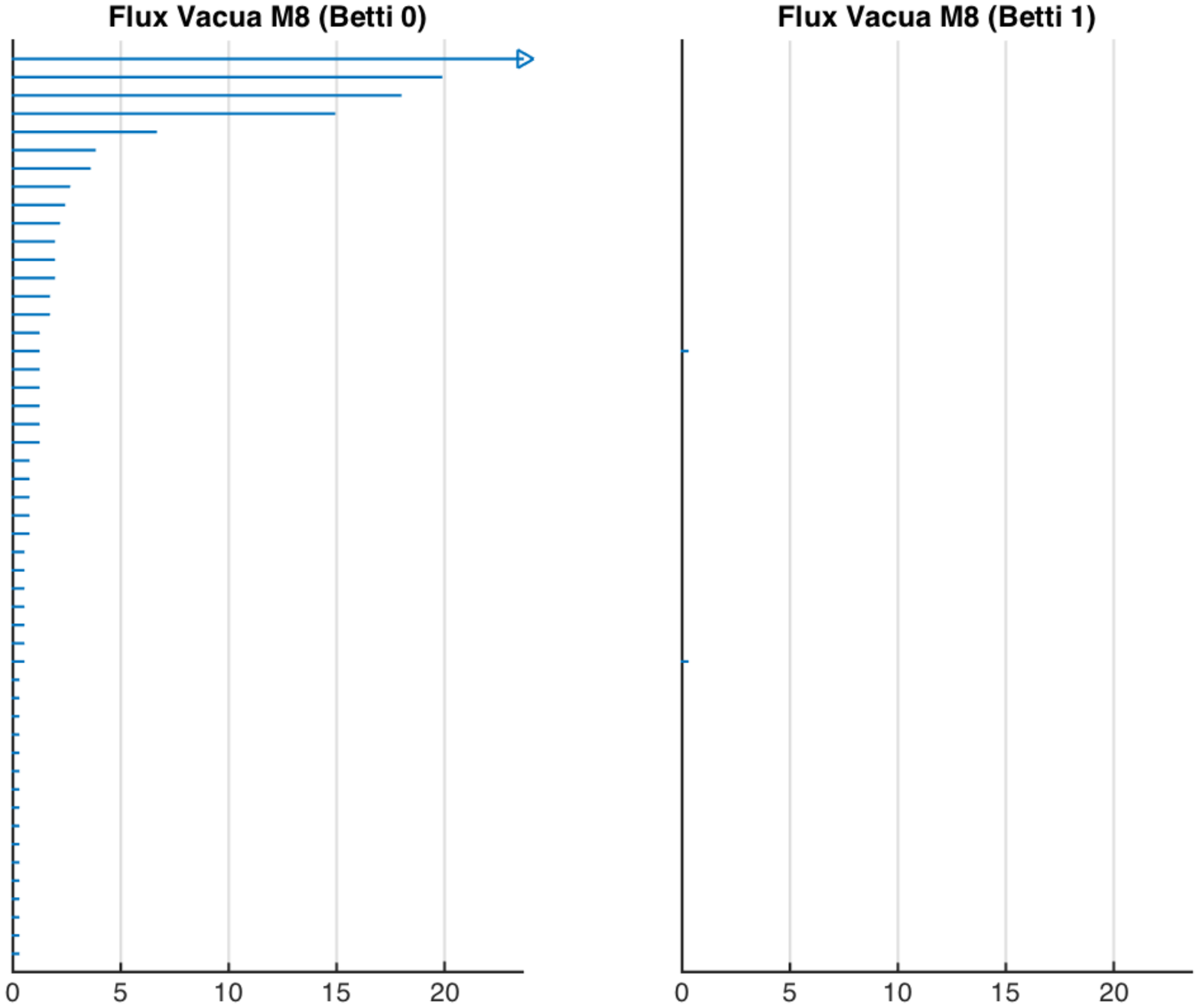}
\caption{Barcodes for the distribution of flux vacua on $X_8 (1^4,4)$. The point cloud $\mathsf{X}$ parametrizes 14585 vacua with vectors of the form $\mathsf{x} = (\re x  , \im x , \re \tau , \im \tau)$. In the persistent homology computation we use the lazy witness complex $\mathsf{LW} (\mathsf{X} , \mathsf{L} , \epsilon)$ with a landmark set $\mathsf{L}$ consisting of 200 points selected using the \emph{min-max} algorithm. The computation runs over 1099256 simplices.}
\label{FluxVacuaM8}
\end{figure} 
We define a point cloud $\mathsf{X}$ where each point represents a vacuum, a solution given by \eqref{EOMtauM8} and \eqref{EOMcomplM8} and parametrized by vectors of the form $\mathsf{x} = (\re x  , \im x , \re \tau , \im \tau)$. Note that contrary to the previous case, we cannot plot the whole point cloud $\mathsf{X}$, but only its projection onto special planes. The persistent homology of $\mathsf{X}$ provides a new way to look at the full set of physical values of the moduli without any projection. Before analyzing the barcodes, we have however to deal with the $\rm{SL} (2 , \zed)$ symmetry acting on the axion-dilaton and on the flux vectors. Its action is
\begin{equation}
\tau \longrightarrow \frac{a \, \tau + b}{c \, \tau + d} \ , \ \qquad \left( \begin{matrix} F_3 \\ H_3 \end{matrix} \right) \longrightarrow  \left( \begin{matrix} a & b \\ c & d  \end{matrix} \right)   \left( \begin{matrix} F_3 \\ H_3 \end{matrix} \right) \, , \ \qquad   \left( \begin{matrix} a & b \\ c & d \end{matrix} \right) \in {\rm SL(2 , \complex) }\ .
\end{equation}
To deal with this redundancy we proceed as follows. First we compute \eqref{EOMtauM8} for a randomly generated set of fluxes ${\mbf f} = (f_1,f_2,f_3,f_4)$ and ${\mbf h}=(h_1,h_2,h_3,h_4)$ using \textsc{mathematica}. Then we map each value of $\tau$ to its fundamental domain (and transform the flux vectors accordingly). To do so we use a standard algorithm, by iterating the transformations $\tau \longrightarrow \tau  - \rm{Round} (\re \tau)$ and $\tau \longrightarrow - \frac{1}{\tau}$ until we reach the fundamental domain\footnote{In practice, to reduce the computation time we put a cutoff to the number of iterations, and simply discard the point if it has not reached the fundamental domain by the time the cutoff is reached.}. Then we use the value of $\tau$ in the fundamental domain and the value of the transformed fluxes to compute $\log x$ from \eqref{EOMcomplM8}. To fix the conifold monodromy we only consider points such that $\arg x \in [ - \pi , \pi )$, and because of the validity range of the approximated periods, we only keep points with $x < 1$.

We have computed the homology of the point cloud using the lazy witness complex $\mathsf{LW} (\mathsf{X} , \mathsf{L} , \epsilon)$ and the barcodes are shown in Figure \ref{FluxVacuaM8}. The only non trivial barcode is for Betti number zero, higher persistent homology groups showing scant or no structure. In this case the dataset does not present any appreciable ``voids". This is due to the tendency of flux vacua to cluster around the conifold singularity. Indeed this clustering shows up in Figure \ref{FluxVacuaM8} as the fact that generic bars are very short-lived. 
\begin{figure}[h]
\centering
\def\svgwidth{0.4cm}
\includegraphics[width=0.7\textwidth]{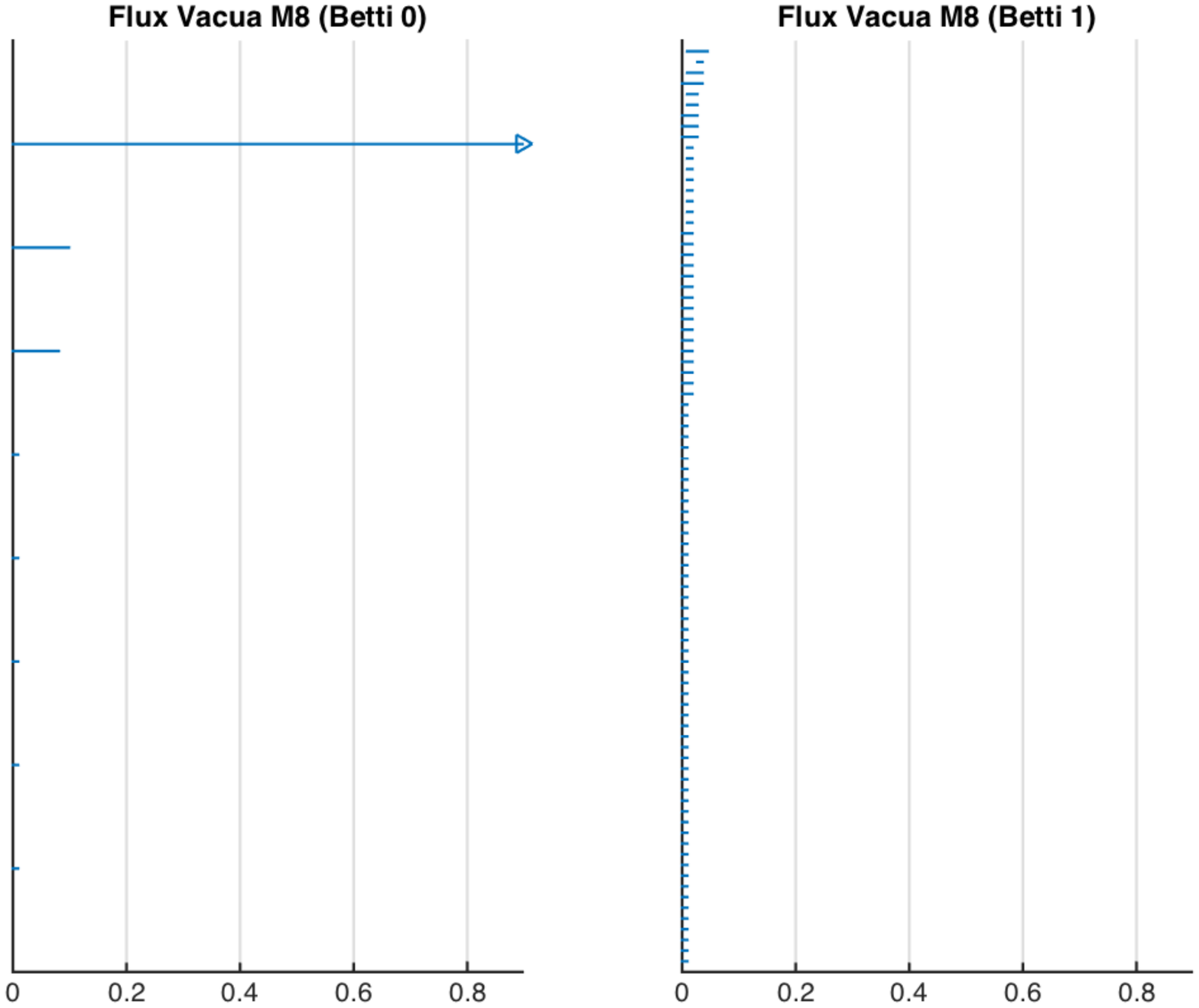}
\caption{Barcodes for the distribution of flux vacua on $X_8 (1^4,4)$. The point cloud $\mathsf{X}$ is constructed as in Figure \ref{FluxVacuaM8}. Also in this case the persistent homology computation is done using the lazy witness complex $\mathsf{LW} (\mathsf{X} , \mathsf{L} , \epsilon)$, but this time the landmark set $\mathsf{L}$ consists of 200 points randomly selected. The computation requires 852691 simplices.}
\label{FluxVacuaM8R}
\end{figure} 
Figure \ref{FluxVacuaM8} also shows three bars which are rather long-lived, with respect to the others (a fourth bar corresponds to the overall connected component at large $\epsilon$). Isn't this a contradiction with the clustering of vacua? The answer is no, and the reason is that we must also be careful in understanding the features of the approximation scheme we are using for our computation. Indeed in the lazy witness scheme we have chosen the landmark selector with the \textit{min-max} algorithm, which selects points spread apart as much as possible. Therefore points which are outside the clustering region will be privileged with respect to points inside, and this is the reason we see these extra bars. These bars correspond to points outside the clustering region, and since this is relatively empty the bars are longer than the average.

To clarify this point further, let us repeat the persistent homology computation, this time using a random selector to choose the landmark set $\mathsf{L}$. Since the selector is random, from the clustering hypothesis we expect that almost all the landmark points will be in the clustering region, and therefore the barcodes' plot will show almost no structure. Indeed this is what we see in Figure \ref{FluxVacuaM8R}; in particular we note the different range of the proximity parameter $\epsilon$ respect to Figure \ref{FluxVacuaM8}.

This example shows which kind of information we can get from the persistent homology of a point cloud of vacua, but also how different approximation schemes have to be handled with care.

\section{A first look at heterotic models} \label{heterotic}

Finally we conclude with a rather preliminary glance at a collection of phenomenologically interesting heterotic vacua. We will consider compactifications of the $\rm E_8 \times E_8$ heterotic string over a Calabi-Yau $X$. The heterotic string has very appealing features in the construction of phenomenological models of physics beyond the standard model \cite{Candelas:1985en}. 

Let us quickly review $\rm E_8 \times E_8$ heterotic $\cN=1$ vacua. The data required for the compactification on $X$ are two holomorphic vector bundles $\cV$ and $\tilde{\cV}$ (or more generically coherent sheaves) with structure group a subgroup of $\rm E_8$ and a number of NS5 branes wrapping a holomorphic curve $C$ in $X$. The bundle $\cV$ is then used to construct a standard model-like effective action while $\tilde{\cV}$ corresponds to the hidden sector. The low energy effective action is $\cN=1$ supergravity coupled to a number of gauge and matter multiplets. Consistency with the low energy Bianchi identity requires a cohomological condition relating $\cV$, $\tilde{\cV}$, $C$ and $X$. Typically one considers $\tilde{\cV}$ to be trivial and the number of NS5 branes to be an adjustable parameter so that the only non-trivial condition involves the visible bundle $\cV$ and $X$. Solving this condition determines a viable structure group $H$ for $\cV$. The low energy \textsc{gut} theory is then based on a group $G$, the commutant of $H$ in $\rm E_8$.

The simplest choice is the standard embedding, which identifies $H$ with the $\rm SU(3)$ holonomy group of $X$ giving an $\rm E_6$ \textsc{gut}. More general non-standard embeddings are possible as long as $\cV$ obeys a stability condition. To every holomorphic bundle or sheaf $\cV$ we associate its slope as the ratio between its degree and its rank $\mu (\cV) = \frac{\deg \cV}{\rank \, \cV}$. Then a bundle $\cV$ is $\mu$-stable if for any sub-bundle $\mathcal{J} \subset \cV$ with $0 < \rank \, \mathcal{J} < \rank \, \cV$ we have $\mu(\mathcal{J}) < \mu (\cV)$. A poly-stable bundle is a direct sum of stable bundles, all of them with the same $\mu$. A poly-stable holomorphic bundle $\cV$ corresponds to a solution of the Donaldson-Uhlenbeck-Yau equations and therefore preserves supersymmetry.

The compactifications that we will consider rely on the choice of a poly-stable vector bundle $\cV$ on $X$. The choice of this bundle breaks the $\rm E_8$ gauge symmetry of the visible sector down to a \textsc{gut} group $G$, for example $\rm SU(5)$ or $\rm SO(10)$, the commutant of the structure group $H$ of $\cV$ in $\rm E_8$. A similar setup holds for the $\rm E_8$ hidden sector. By turning on appropriate Wilson lines it is possible to further break $G$ to the standard model group plus a number of $U(1)$ factors. To have non-trivial Wilson lines $X$ has to be non-simply connected, which is in general not the case for Calabi-Yaus constructed from complete intersections in weighted projective spaces. One can easily remedy this problem by considering a discrete group $\Gamma$ freely acting on $X$ and then working equivariantly with respect to $\Gamma$. In practice this is accomplished by working with the Calabi-Yau $\hat{X}= X / \Gamma$, for which $\pi_1 (\hat{X}) = \Gamma$. Each element $\gamma \in \Gamma$ corresponds to a 1-cycle. Wilson line operators $W_{\gamma}$ can be defined along these cycles in terms of a flat rank one bundle on $\hat{X}$. The choice of a bundle $\cV$ equivariant with respect to the $\Gamma$-action descends to a bundle $\hat{\cV}$ on $\hat{X}$. The physical spectrum and the relevant low energy quantities can then be computed from the data of $X$ and $\cV$. Physically the Wilson line operators break the \textsc{gut} group $G$ to the subgroup commuting with all the Wilson lines. The low energy effective field theory on $\hat{X}$ is based on said group and can be chosen to be a phenomenologically realistic model. 

In an impressive series of works \cite{Anderson:2011ns,Anderson:2012yf,Anderson:2013xka}, a large class of realistic models were constructed along these lines, containing the standard model gauge group, the matter content of the MSSM and no exotics. With an extensive use of computational algebraic geometry many low energy properties of such models were derived explicitly and are available in the database \cite{HetLine}.

If we choose $\cV$ to have a rank five special unitary structure group, so that  the first Chern class $c_1 (\cV)$ vanishes, the \textsc{gut} group will be $\rm SU(5)$, up to abelian factors. The latter are typically Green-Schwarz anomalous and decouple at high energies. The requirement of low energy supersymmetry implies that $\cV$ has to be a $\mu$-polystable bundle, a direct sum of $\mu$-stable bundles all with the same $\mu$-stability parameter. A clever choice is a sum of line bundles
\begin{equation} \label{Vsplit}
\cV = \bigoplus_{a=1}^5 \, \cL_a \, ,
\end{equation}
with $\mu (\cL_a) = 0$ for all $a$. It is quite remarkable that such a simple choice still allows for physically realistic models, and indeed a scan over such possibilities (and other closely related) done in \cite{Anderson:2011ns,Anderson:2012yf,Anderson:2013xka} revealed a large class of viable models. 

We want to have a preliminary look at this database from the perspective of persistent homology. In particular we will only consider varieties $X$ which are CICYs with $\cV$ of the form \eqref{Vsplit}, despite other possibilities being allowed. We want to construct a point cloud $\sfX$ out of this dataset, and we will do so in the simplest possible way, by collecting vectors of the form $\mathsf{x} = \left( h^{1,1} (X) , h^{2,1} (X) , c_2 (\cL_1) , \dots,c_2 (\cL_5) \right)$. This simple choice certainly does not do proper justice to the database built in \cite{Anderson:2011ns,Anderson:2012yf,Anderson:2013xka} which contain extensive geometrical and physical informations on the models. In particular such a parametrization is very coarse since a point in $\sfX$ represents many physically distinct models. Such a problem can be resolved by adding more and more parameters in the construction of the point cloud $\sfX$, but we will leave such a detailed analysis for the future.

On the other hand such a point cloud $\sfX$ is embedded in $\real^7$: it cannot be plotted in any simple way and there is no readily available tool to gauge the properties of this distribution of vacua. We propose that persistent homology provides such a tool.
\begin{figure}[htbp]
\centering
\def\svgwidth{0.4cm}
\includegraphics[width=0.8\textwidth]{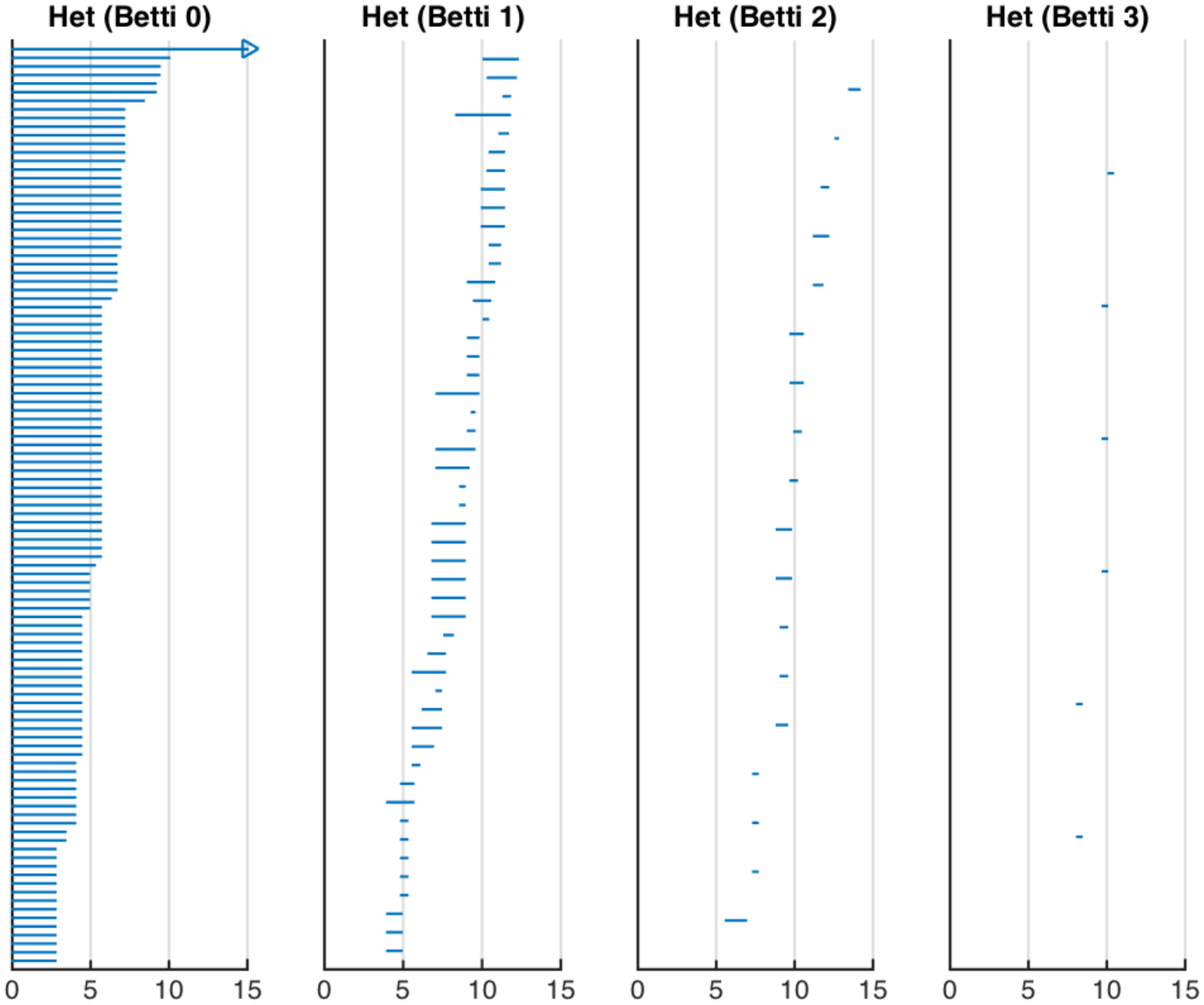}
\caption{Barcodes for the distribution of heterotic vacua. The datum of a vacuum consists in the two Hodge numbers $h^{1,1} (X)$ , $h^{2,1} (X)$ of the Calabi-Yau and the second Chern classes $c_2 (\cL_i)$ of five line bundles, such that $\cV = \bigoplus_{a=1}^5 \, \cL_a$ specifies the \textsc{gut} spectrum. The point cloud $\mathsf{X}$ consists of 107 points in $\real^7$. The homological computation is done using the Vietoris-Rips complex $\mathsf{VR}_\epsilon (\mathsf{X})$ and runs over 1786924 simplices.
} 
\label{H11H21fiveC2}
\end{figure} 
We have considered around a hundred of models with distinct values of $\mathsf{x} \in \sfX$ and computed the $\IN$-persistence modules given by the homology of the filtered Vietoris-Rips complex $\mathsf{VR}_{\epsilon} (\sfX)$. We show the barcodes in Figure \ref{H11H21fiveC2}. The barcodes show plenty of structure, even in degree four, if very short-lived.

Figure \ref{H11H21fiveC2} is a concrete, qualitative example of one of the main ideas of this note: that distributions of physically interesting vacua show a certain degree of topological complexity. In the distribution at hand we see the appearance of many $n$-cycles. This could imply for example that there exists subtle correlations between the vacua, in the form of some algebraic equations which describes the $n$-cycle. Of course to establish this decisively and quantitatively one has to go beyond the topological analysis presented here. 

On the other hand the topological analysis is very efficient in comparing different distributions of vacua. As we have explained our working assumption is that distributions of physically realistic models are not random but have topological features. In the light of this idea, one could start comparing different distributions of vacua and look for the ones which exhibit more structure. Distribution with higher complexity are, in a certain sense, singled out.

Let us try to explore this possibility more in detail and compare directly two distributions of vacua by fixing certain parameters. Out of the database of \cite{HetLine} we can form more refined point clouds with more information. For example an information readily available is the number $n_{\mathbf{5}}$ of vector-like pairs $\mathbf{5}$ and $\mathbf{\bar{5}}$ at the \textsc{gut} level, or the number of physical Higgs doublets $n_{\mathbf{H}}$ in the low energy spectrum. We take these two parameters as inputs together with $\sfX$ to create an augmented point cloud $\mathsf{Y}$. Each point $\mathsf{y} \in \mathsf{Y}$ contains the same geometrical information as above plus this extra information on the spectrum and has the form $\mathsf{y} =  \left( h^{1,1} (X) , h^{2,1} (X) , c_2 (\cL_1) , \dots,c_2 (\cL_5) , n_{\mathbf{5}} , n_{\mathbf{H}} \right) $. However this time, instead of just studying the topological features of $\mathsf{Y}$, we select two subsets of data $\mathsf{Y}_1$ and $\mathsf{Y}_3$ as follows. All the points in both subsets have a single Higgs doublet $ n_{\mathbf{H}} =1$, but they differ in $n_{\mathbf{5}}=1$ and $n_{\mathbf{5}}=3$ respectively\footnote{These precise values have been chosen just for convenience.}. 
\begin{figure}[htbp]
\centering
\def\svgwidth{0.4cm}
\includegraphics[width=0.8\textwidth]{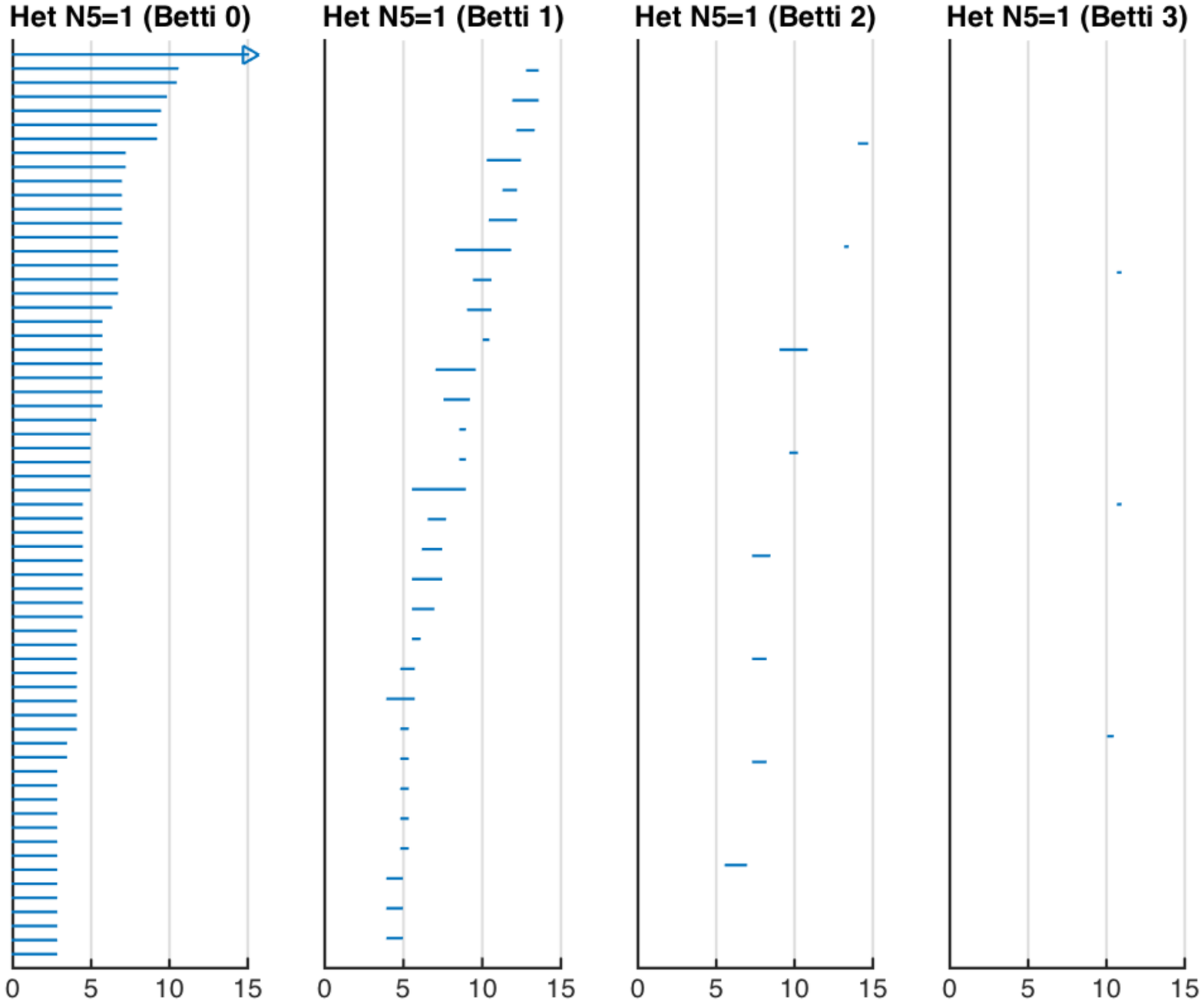}
\caption{Barcodes for the point cloud $\mathsf{Y}_1$ consisting of 65 points in  $\real^9$ which parametrize an heterotic compactification as in Figure \ref{H11H21fiveC2}, with two additional parameters, the number of Higgs doublets $n_{\mathbf{H}} =1$ and of vector-like pairs  $n_{\mathbf{5}}=1$. The computation of the persistent modules $H_i (\mathsf{VR_\epsilon (\mathsf{Y}_1)} ; \zed_2)$ involves 574095 simplices.}
\label{1Higgs1Num5}
\end{figure} 
\begin{figure}[htbp]
\centering
\def\svgwidth{0.4cm}
\includegraphics[width=0.8\textwidth]{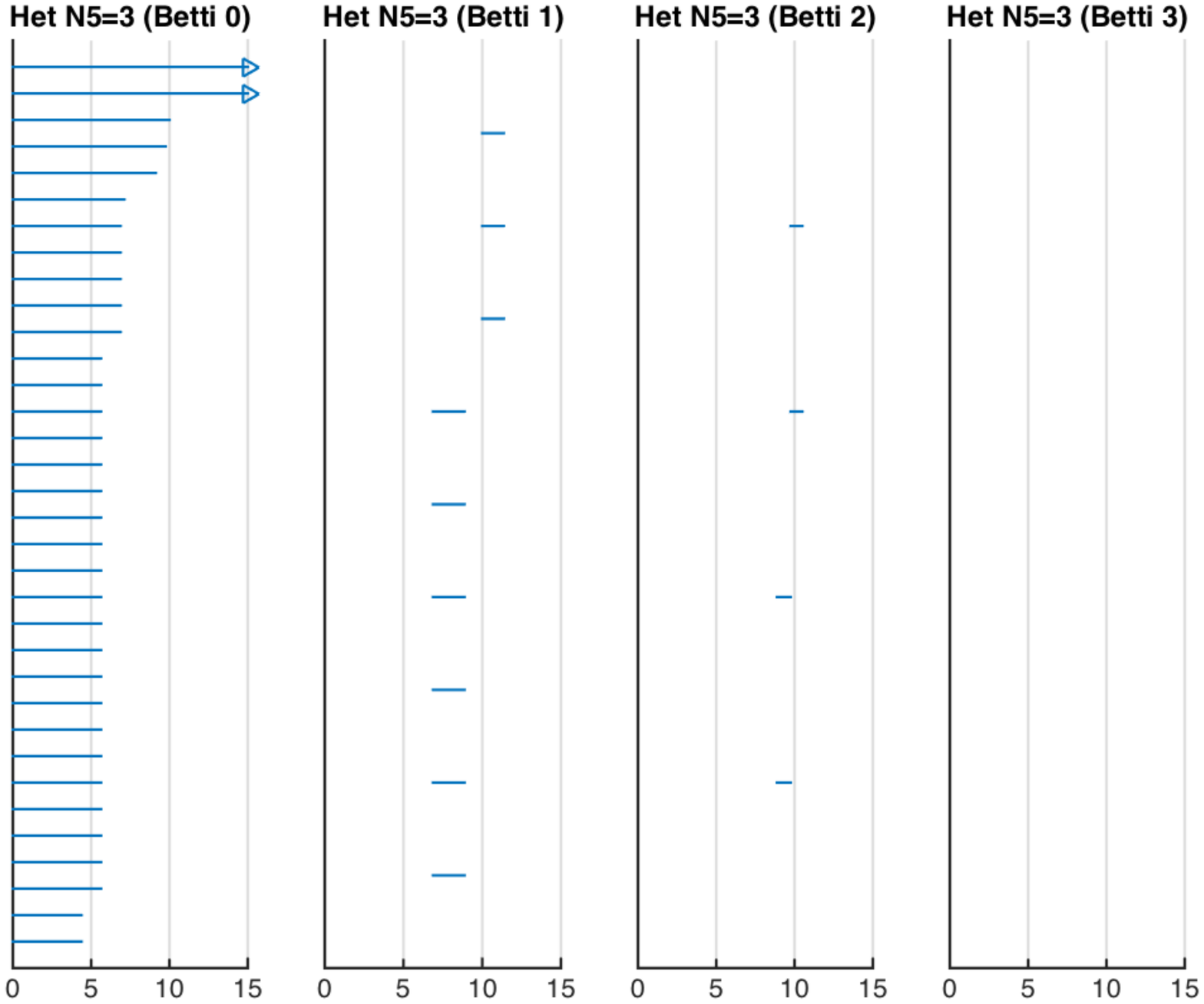}
\caption{Barcodes for the point cloud $\mathsf{Y}_3$ consisting of 34 points in $\real^9$ as in Figure \ref{1Higgs1Num5}, the only difference being the value of vector pairs which is now fixed at $n_{\mathbf{5}}=3$. The persistent homology computations runs over 10325 simplices}
\label{1Higgs3Num5}
\end{figure} 
In other words the datasets of $\mathsf{Y}_1$ and $\mathsf{Y}_3$ contain geometrical parameters of a Calabi-Yau and an holomorphic bundle which give rise to different spectra with fixed characteristics. We want to compare their distributions from the perspective of their persistent homology. Again we study the filtered Vietoris-Rips complexes $\mathsf{VR}_{\epsilon} (\mathsf{Y}_1)$ and $\mathsf{VR}_{\epsilon} (\mathsf{Y}_3)$ and collect the information about their $\IN$-persistence modules in the barcodes shown in Figures \ref{1Higgs1Num5} and \ref{1Higgs3Num5}.

Although comparisons have to be made carefully since the two datasets have different number of entries, one feature is immediately clear: among the database of \cite{HetLine} for models with a single Higgs doublet, the distribution of vacua with $n_\mathbf{5}=1$ is topologically much more complex than that of  $n_\mathbf{5}=3$ vacua. The latter indeed shows clear regularities in the barcodes. This is clear by comparing Figures \ref{1Higgs1Num5} and \ref{1Higgs3Num5}, but is also from the number of simplices generated during the persistent homology computation, which is greater for $\mathsf{Y}_1$ by a factor of 50.

If we assume that topologically rich structures correspond to physically interesting features, one would conclude that compactifications with $n_\mathbf{5}=1$ would be preferred over compactifications with $n_\mathbf{5}=3$. Again, this lends (modest) support to the idea that string vacua can be characterized by the topological properties of their distributions. In particular it opens up the possibility that vacua with realistic features can be singled out by the complexity of their persistent homology. In other words we are led to propose a qualitative vacuum selection principle based on the topological features of the distributions of vacua:  topologically complex distributions of vacua are preferred over topologically simple ones. We believe this idea is worth further investigations.

\section{Conclusions} \label{conclusions}

In this paper we have explored the possibility that distributions of string vacua can be characterized by certain topological properties. A readily available tool to capture the topological features of a distribution of points is topological data analysis. In this framework persistent homology extracts topological invariants at every length scale. To a collection of vacua we associate persistence modules in various degree; the latter in turn are completely characterized by their barcodes. Barcodes are a graphical representation of the lifespan of the persistent homology classes of the distribution of vacua, as the length scale at which we look at the distribution is varied.

To exemplify these ideas we have studied three classes of vacua. In Section \ref{N2vacua} we have studied vacua with $\cN=2$ supersymmetry, arising from  compactifications of the type II string on a Calabi-Yau, or from Landau-Ginzburg models. By appropriately labeling families of such vacua we have constructed the corresponding point clouds and computed their persistent homologies. We have looked for topological structures in the distribution of certain Calabi-Yau manifolds, or superconformal fixed points. Both of these are interesting mathematical problems on their own, regardless of the physical applications. During the process we have learned how to use topological data analysis to reproduce the known features of such distributions. We have also established that certain distributions of vacua, such as Calabi-Yau with small Hodge numbers, have non-trivial topological features. Note that, regardless of the interpretation as string vacua, the study of Landau-Ginzburg distributions could also be seen as a very preliminary attempt at the study of the topology of the space of two dimensional superconformal field theories. It would be very interesting to understand if persistent homology could be an useful tool for this direction of investigation.

In Section \ref{flux} we have applied our formalism to certain classes of flux vacua in type II\,B/F-theory compactifications. The models we have discussed are typically presented as examples of the statistical approach to the landscape of flux vacua, where one is interested in counting the number of vacua with certain properties. In this framework persistent homology has the advantage of allowing a simple visualization of the basic properties of a distribution of vacua, even when such a distribution has high dimension. We have discussed how the features of such distributions can be extracted from their barcodes in two simple cases. We have given concrete examples of how the techniques of topological data analysis can be used to determine properties of the distributions of flux vacua, not just by counting their numbers but also by using the invariants of persistent topology.

In this note we have obtained distributions of flux vacua by solving directly the equations for supersymmetric vacua. Of course another possibility would be to use directly a continuum approximation to the index density of vacua as in \cite{Douglas:2003um,Ashok:2003gk,Douglas:2005df,Denef:2004ze}. One can easily construct a point cloud by discretizing an index density such as \eqref{indexvacua} and study its properties using statistical topology. Indeed we expect that this should be a rich area of investigation. 

Finally in Section \ref{heterotic} we have considered quasi-realistic compactifications of the heterotic string. We have considered a class of vacua labeled by a Calabi-Yau and the topological data of a certain holomorphic bundle. A striking aspect of such a class of vacua is that it exhibits non-trivial topological features, in the form of higher dimensional, if short-lived, $n$-cycles. The presence of such features suggests that distributions of phenomenologically viable vacua might be characterized by a high degree of topological complexity, as seen from their persistent homology modules. In other words it leads to the possibility that physically interesting features are associated with topologically interesting structures at the level of persistent homology. This possibility can be made rather concrete by comparing the persistent topological features of different classes of vacua. We have given a specific example by comparing two different set of vacua which differ in the number of $\mathbf{5}-\mathbf{\bar{5}}$ pairs in the \textsc{gut} spectrum. The class of vacua with $n_{\mathbf{5}}=1$ is topologically much richer and therefore from our perspective should be preferred. 

Let us summarize the main ideas we have encountered in this note. First of all we have shown how techniques of topological data analysis can be applied to the problem of characterizing distributions of string vacua. Persistent homology can be used to extract qualitative informations from the set of string vacua. The usefulness of such informations depends on the questions one is asking. Certain features of string vacua can be understood from the more persistent homology classes in the barcodes; others cannot. Hopefully such techniques could be efficiently applied in conjunction with the usual tools from statistics and algebraic geometry typically used in studying distributions of vacua. We also hope to call the attention of the computational topology community on the extremely rich and diversified problem posed by the landscape of string vacua.

A particular interesting point in this respect is that, also due to the approximation schemes available, studying the persistent topology of a distribution of vacua is considerably easier than studying its geometrical properties. In other words to sample the relevant features one typically needs only a smaller number of vacua and the whole process is computationally less involved. On the other hand the usual ideas used in the study of persistent homology need to be partly revisited to be applied in this context: we have seen an example of very short-lived bars which reveal physically interesting features only when regarded in correlation with barcodes in other degrees. To which extent topological data analysis is physically relevant is at the moment not clear; certainly more work and ideas are needed to understand how much information and how many new insights can we gain from it. One could hope to learn something by analyzing much bigger datasets, or by studying systematically families of M- or F-theory compactifications. 

On a more speculative level the results of this note have led us to the idea that sets of vacua with a higher degree of topological complexity are singled out with respect to topologically simpler ones. It is tempting to suggest that the presence of topologically interesting features within a distribution of vacua is associated with physically interesting features. As we have already remarked, the presence of higher dimensional cycles can imply subtle correlations between the vacua distributed along the cycle, although a more refined case by case analysis is needed to establish this concretely. These correlations can assume the form of a set of equations which force the vacua to lie over a certain $n$-cycle. If they exist, such correlations would distinguish a family of vacua from others. Even if at this stage we have no general understanding of the physical significance of such correlations (except when they arise from a superpotential, where in principle it should be possible to derive them analytically), persistent homology provides a computational framework to visualize them. This leads us to the possibility of formulating a qualitative topological vacuum selection principle: that \emph{topologically complex distributions of vacua are physically preferred over topologically simple ones}. Here the concept of topological complexity is concretely provided by persistent homology. Clearly much more work is required to turn this into a quantitative statement.

%For example within the context of flux compactifications, a high degree of complexity of the bundle in which the superpotential takes values implies a larger number of vacua. Such portions of the landscape are more populated.

\section*{Acknowledgements}

I wish to thank the Applied Topology group at Stanford University for making \textsc{javaplex} available in \cite{javaplex}. Similar thanks are in order to Andre Lukas and Harald Skarke for maintaining the databases used in this note and keeping them available for use. This work was partially supported by FCT/Portugal and IST-ID through UID/MAT/04459/2013, EXCL/MAT-GEO/0222/2012 and the program Investigador FCT IF2014, under contract IF/01426/2014/CP1214/CT0001. I also thank IHES for hospitality and support during the preparation of this paper.

\end{document}